\documentclass[smallextended]{svjour3}
\smartqed  

\usepackage{hyperref}

\usepackage{amssymb, amsmath, latexsym, enumitem, makeidx, epsfig, url,hyperref}
\usepackage{pdfpages}
\usepackage{graphicx}
\usepackage{array}

\newcommand{\squeezeup}{\vspace{-5cm}}

\newcolumntype{C}[1]{>{\centering\arraybackslash}p{#1}}

\reversemarginpar
\begin{document}

\title{On Higher Order Query Languages which on Relational Databases Collapse to Second Order Logic
\thanks{The research reported in this paper results from the project {\it Higher-Order Logics and Structures} supported by the {\bf Austrian Science Fund (FWF):[I2420-N31]}. This work was initiated during the research visit of Professor Jos\'{e} Mar\'{i}a Turull-Torres in the frame of the project {\it Behavioural Theory and Logics for Distributed Adaptive Systems} supported by the {\bf Austrian Science Fund (FWF):[P26452-N15]}.}}\author{Flavio Ferrarotti \and Loredana Tec \and Jos\'{e} Mar\'{i}a Turull-Torres}
\titlerunning{On Higher Order Query Languages} 


\institute{Flavio Ferrarotti \at
Software Competence Center Hagenberg, Austria,\\
  \email{Flavio.Ferrarotti@scch.at}
  \and
Loredana Tec \at 
Software Competence Center Hagenberg, Austria,\\
  \email{Loredana.Tec@scch.at}
\and 
Jos\'{e} Mar\'{i}a Turull-Torres
\at
Depto. de Ingenier\'ia e Investigaciones Tecnol\'ogicas
Universidad Nacional de La Matanza, Argentina,
Massey University, New Zealand,\\
 \email{J.M.Turull@massey.ac.nz}}



\date{Submitted \today}

\maketitle

\begin{abstract}
In the framework of computable queries in Database Theory, there are many examples of queries to (properties of) relational database instances that can be expressed by simple and elegant third order logic ($\mathrm{TO}$) formulae. In many of those properties the expressive power of $\mathrm{TO}$ is not required, but the equivalent second order logic ($\mathrm{SO}$) formulae can be very complicated or unintuitive. From the point of view of the study of highly expressive query languages, it is then relevant to identify fragments of $\mathrm{TO}$ (and, in general, of higher-order logics of order $\geq 3$) which \textit{do have} an $\mathrm{SO}$ equivalent formula. In this article we investigate this precise problem as follows. Firstly, we define a general \textit{schema} of $\exists\mathrm{TO}$ formulae which consists of existentially quantifying a third order linear digraph of polynomial length, that is, a sequence of structures that represents a computation, by explicitly stating which operations are the ones which can be involved in the construction of a given structure in the sequence, when applied to the previous one. Then we give a constructive proof of the fact that all $\exists\mathrm{TO}$ subformulae of that schema can be translated into an equivalent $\mathrm{SO}$ formula. We give several examples which show that this is a very usual, intuitive, and convenient schema in the expression of properties. Secondly, aiming to formally characterize the fragment of $\mathrm{TO}$ which can be translated to $\mathrm{SO}$, we define a restriction of $\mathrm{TO}$, which we denote $\mathrm{TO}^{P}$, for \textit{polynomial $\mathrm{TO}$}, and we give a constructive proof of the fact that it collapses to $\mathrm{SO}$. 
We define $\mathrm{TO}^{P}$ as the fragment of $\mathrm{TO}$ where valuations can assign to $\mathrm{TO}$ relation variables only $\mathrm{TO}$ relations whose cardinalities are bounded by a polynomial that depends on the quantifier. Moreover, we define a similar restriction for every higher order logic of order $i \geq 4$ ($\mathrm{HO}^i$), which we denote $\mathrm{HO}^{i,P}$, for \textit{polynomial $\mathrm{HO}^{i}$}, and we give a constructive proof of the fact that for all $i \geq 4$, $\mathrm{HO}^{i,P}$ collapses to SO.
\end{abstract}

\section{Introduction}
\label{sec:examples_HO}

In the framework of computable queries to relational databases and Finite Model Theory where queries define (second order) relations on the input database instance or finite relational structure, there are many examples of properties (queries) that can be expressed by simple and
elegant third order logic (TO) formulae \footnote{TO extends second order logic with third order quantifiers which bind third order relation variables;  those variables are valuated with sets of tuples of (second order) relations.}.
Let us consider three such properties:

\begin{enumerate}[label=\emph{\alph*})]

 \item
Consider the property \emph{hypercube graph}
(see~\cite{[Fer08]}).
An \emph{$n$-hypercube graph} ${\bf Q}_n$, also called
an $n$-cube,  is
an undirected graph whose vertices are binary $n$-tuples.
Two vertices of ${\bf Q}_n$ are adjacent if they differ in exactly one bit.
 Note that we can build an $(n+1)$-cube ${\bf Q}_{n+1}$ starting with two
  isomorphic copies of an $n$-cube ${\bf Q}_n$ and adding edges between
  corresponding vertices.
 Using this fact, we can define in
     TO the class of \textit{hypercube graphs}, by saying that there is a sequence of graphs (i.e., a \emph{third order linear digraph}, where every TO node is an undirected (second order) graph) which starts with the graph $K_2$, ends with a graph which is equal to the input graph, and such that
      every graph $G_2$ in the sequence results from finding two total, injective functions $f_1, f_2$ from the previous graph $G_1$, so that $f_1$ and $f_2$ induce in $G_2$ two isomorphic copies of $G_1$, the images of those functions define a partition in the vertex set of $G_2$, and there is an edge in $G_2$ between the images $f_1(x)$ and $f_2(x)$ of every node $x$ in $G_1$.

\item
  Another definition of \textit{hypercube graphs} that yields a simple (TO) formula is the following. We say that there is a proper non empty subset $V'$ of the
    vertex set $V$ of the input graph $G$, and a (third order) bijective function $f$ from the vertex set of $G$ to the power set of $V'$, such that, for every pair of nodes $x$ and $y$ in $G$, there is an edge between them if and only if $f(x)$ can be obtained from $f(y)$ by adding or removing a single element (note that $V'$ is necessarily of size $\log_2 |V|$).

 \item
 As another example consider the \textit{Formula-Value query}, i.e., given a propositional formula with constants in $\{F,T\}$ decide whether it is true. We can express it with a simple and intuitive TO formula by saying that there is a sequence of word models (which represent formulae)  which starts with the input formula, ends with the formula $T$, and such that every formula $\varphi_2$ in the sequence results from the application to the previous formula $\varphi_1$ of one of the operations of conjunction, disjunction, or negation which is ready to be evaluated (i.e., like in $(T \wedge F)$), or elimination of a pair of redundant parenthesis (i.e., like in $((T))$) to exactly one subformula of $\varphi_1$.

\end{enumerate}

    Actually, the expressive power of third order
logic \textit{is not required} to characterize hypercube graphs, since
they can be recognized in nondeterministic polynomial
time, and by Fagin's theorem~\cite{[Fag74]}, $\exists$SO captures NP.
Thus, there are formulae in existential second order logic
($\exists$SO) which can express this property. Nevertheless,
to define the class of hypercube graphs in
second order logic is certainly more challenging than to define it
in TO (see the two strategies for hypercube graphs in~\cite{[FRT14]}).
Also, we  \textit{do not need} third order
logic to express the Formula-Value query, since it is in DLOGSPACE~\cite{[BM92]}, and hence can be expressed in $\exists$SO, since
DLOGSPACE $\subseteq$ P $\subseteq$  NP  $= \exists$SO.

On the other hand, if we consider the query SATQBF of satisfiable quantified Boolean formulae, we can express it in $\exists$TO, since the problem is PSPACE-complete, and it is a well-known fact that $\exists$TO is powerful enough as to characterize every problem in PSPACE, since it captures NTIME($2^{n^{O(1)}}$)  (see~\cite{[HT06]}).
Note that as PSPACE can be captured by SO extended with a transitive closure operator, and furthermore this logic is widely conjectured to be strictly more expressive than the standard second order logic, the existence of an SO characterization of this problem is unlikely.

Then, it would be very interesting to \textit{distinguish} in some way the TO formulae which \textit{do have} an SO equivalent formula, like in the first three examples above, from the TO formulae which (most likely) do not, as in SATQBF. In the general case, it would mean that for those queries  in the first class we can take advantage of the much higher expressibility and simplicity of TO, and be able to express a query in a more simple and  intuitive way, \textit{though still formal}, but \textit{without} having to pay the price of a higher complexity to evaluate the corresponding formulae. Note that by the results in~\cite{[Fag74]} and~\cite{[HT06]}, $\exists$SO $=$ NTIME($n^{O(1)}$), while $\exists$TO $=$  NTIME($2^{n^{O(1)}}$).

In addition, there are well known problems such as hypercube graph (which as we saw above can also be characterized in $\exists$SO) and SATQBF$_{k}$ (i.e., satisfiability of quantified Boolean formulae with $k$ alternating blocks of quantifiers, which can be characterized in $\Sigma^1_k$, since this problem is $\Sigma^p_k$ complete and  $\Sigma^1_k = \Sigma^p_k$, for all $k \geq 1$, see~\cite{[Sto76]}),  that do not appear to have a straightforward characterization in second order logic, even if we consider the full second order language. In~\cite{[FRT14]} we gave detailed formulae for those properties, and the two sentences turned out to be  complex and several pages long.

From an applied perspective, this indicates that it makes sense to investigate higher order logics and structures in the context of database query languages. Despite the fact that most of the queries commonly used in the industry are in P, the use of higher order quantifiers can potentially simplify the way in which many of those queries are expressed. Think for instance of PERT charts, which are extensively used in Software Engineering in the context of planning and scheduling tasks of project management. Formally, these charts correspond to graphs with edges representing tasks or activities that need to be done, while nodes represent events or milestones. In the case of planning and scheduling the many interrelated tasks in a large and complex project, where for instance, a node  represents a PERT chart itself, the encoding of higher order relations of order $\geq 3$ into SO relations as studied in this paper could be exploited as a normal form to store such type of complex higher order objects into a standard relational database. Furthermore, for querying such a complex PERT chart, it becomes necessary to perform a so-called ``zooming'' in order to retrieve, e.g., a sub activity used in the node of a higher level activity or simply to determine if or which sub activities can be done in parallel. All these kinds of queries can be naturally expressed using the higher order logics studied in this work. The translation of higher order queries of order $\geq 3$ to SO logic as proposed in this paper could then be a first fundamental step to synthesise the resulting SO queries into corresponding efficient queries over the (normalized) relational databases, much in the same style as~\cite{Itzhaky:2010}.

Taking into account these considerations, is then relevant to identify ways to isolate the ``good behaving'' fragments of TO (and, in general, HO$^{i}$, for $i \geq 3$) formulae. In this line we define in Section~\ref{sec:schema} a general \textit{schema} of $\exists$TO formulae which generalizes the examples $(a)$ and $(c)$ above, and we give a constructive proof of the fact that all $\exists$TO subformulae of that schema can be translated into an equivalent SO formula.
The schema is essentially the expression of an iteration of polynomial length, unfolded as a sequence of relational structures
which represents a computation or derivation in the sense of Complexity and Computability Theories, by explicitly stating which operations are the ones which can be involved in the construction of a given structure in the sequence, when applied to the previous one. This is a very usual, intuitive, and convenient schema in the expression of properties (see Section~\ref{sec:schema} for further examples). 

Then, in Section \ref{sec:top}  aiming to formally characterize the fragment of TO which can be translated to SO, we define  a restriction of TO, which we denote TO$^{P}$, for \textit{polynomial TO}, and we give a constructive proof of the fact that it collapses to SO. We conjecture that TO$^{P}$ is the exact characterization of the class of TO formulae which can be translated to SO, but we do not think that it can be proved.
We define TO$^{P}$ as the fragment of TO where  valuations can assign to TO relation variables only  TO relations whose cardinalities are bounded by a polynomial that depends on the quantifier. Note that the example $(b)$ above does not seem to be expressible by a TO formula of the \textit{well-behaved} schema described above, but is clearly expressible in $\exists$TO$^{P}$. Nevertheless, in the final part of this section we argue that the general schema of $\exists$TO formulae proposed in Section~\ref{sec:schema} is, from the perspective of database query languages, still relevant. Regarding expressive power, we also discuss briefly (in the light of our result) the case of SO extended with the deterministic inflationary fixed-point (IFP) quantifier, where the variable which is bounded by the IFP quantifier is a third order variable. If this SO+IFP logic is restricted to fixed points with a polynomially bounded number of stages, then it also collapses to SO. Moreover, we define a hierarchy of prenex TO$^P$ formulae and show its exact correspondence with the polynomial time hierarchy.        


Finally, in Section~\ref{sec:HOip}, we define a similar restriction of all higher order logics HO$^{i}$, for each $i \geq 4$, which we denote HO$^{i,P}$, for \textit{polynomial HO$^{i}$}. Then we give a constructive proof of the fact that for all $i \geq 4$, $\mathrm{HO}^{i,P}$ collapses to SO.
Roughly, HO$^{i}$ is first order logic extended with quantifiers of any order $2 \leq j \leq i$, which in turn bind $j$-th order relation variables. For $j \geq 3$,  $j$-th order relation variables are valuated with sets of tuples of $(j-1)$-th order relations. A second order relation is a relation in the usual sense (i.e., fixing  $r \geq 1$, a set of $r$-tuples of elements from the domain of a structure), and for $j \geq 3$, a $j$-th order relation is a set of tuples of $(j-1)$-th order relations, of some fixed width.
For HO$^{i,P}$ we use a different strategy which we give in detail for the case of the collapse of HO$^{4,P}$ to SO (see also Appendices~\ref{sectionA} and ~\ref{sectionB}).  That strategy can be generalized in a straightforward way to all orders $i \geq 4$. To define HO$^{i,P}$ we do it inductively, starting with HO$^{4,P}$, stating that HO$^{i+1,P}$ is an extension of HO$^{i,P}$, where the $(i+1)$-th order quantifiers restrict the cardinality (i.e., the number of tuples of $(i)$-th relations)  of the valuating $(i+1)$-th order relations to be  bounded by a polynomial that depends on the quantifier.

\section{Preliminaries}
We assume the reader is familiar with the basic concepts and the framework of Finite Model Theory~\cite{[EF99]}. We only consider signatures, or vocabularies, which are purely \emph{relational}, and for simplicity we do not allow constant
symbols. We use the classical Tarski's semantics, except that in the context of finite model theory, \emph{only finite} structures or interpretations are considered. 
Thus our structures will always be \emph{finite relational structures}. If ${\bf I}$ is a structure of vocabulary $\sigma$, or $\sigma$-structure, we denote its domain by $\textit{dom}(\mathbf{I})$ or $I$, which is a finite set containing all elements of the structure.
Recall that a \emph{valuation} is a function which assigns to every variable in the logic, an element of a given structure. By $\varphi(x_1,\ldots,x_r)$ we denote a formula of some logic whose free variables are exactly $\{x_1,\ldots,x_r\}$. If $\varphi(x_1,\ldots,x_r)$ is a formula of vocabulary  $\sigma$, ${\bf I}$ is a $\sigma$-structure, and $\bar{a}=(a_1,\ldots,a_r)$ is an $r$-tuple over $I$, we use the notation ${\bf I},v \models \varphi(x_1,\ldots,x_r)[\bar{a}]$ to denote that $\varphi$ is satisfied by the structure ${\bf I}$ under the valuation $v$ and that $v(x_i)=a_i$ for $1 \le i \le r$. In turn, the notation ${\bf I} \models \varphi(x_1,\ldots,x_r)[\bar{a}]$ denotes that $\varphi$ is satisfied by the structure ${\bf I}$ under \emph{all} valuations $v$ such that $v(x_i)=a_i$ for $1 \le i \le r$. 

With HO$^{i}$,  for any $i \geq 2$, we denote $i$-th order logic which extends first order logic with quantifiers of any order $2 \leq j \leq i$, which in turn bind $j$-th order relation variables. In particular, HO$^2$ denotes second order logic (SO) as usually studied in the context of finite model theory (see \cite{[EF99],[Lib04]} for a formal definition), and HO$^3$ denotes third order logic (TO). 
Second order variables of arity $r$ are valuated with $r$-ary relations in the usual sense (i.e., a set of $r$-tuples of elements from the domain of a structure). 
For $j \geq 3$,  $j$-th order relation variables are valuated with sets of tuples of $(j-1)$-th order relations according to their relation types. 
A \textit{third order relation type of width} $w$ is a $w$-tuple $\tau = (r_1,\ldots, r_{w})$ where  $w, r_1,\ldots, r_{w} \geq 1$, and $r_1,\ldots, r_{w}$ are arities of (second order) relations.
For $i \geq 4$, an \textit{$i$-th order relation type of width} $w$ is a $w$-tuple $\tau =
(\rho_1,\ldots, \rho_{w})$ where  $w \geq 1$ and $\rho_1,\ldots, \rho_{w}$ are $(i - 1)$-th order relation types.
A \textit{second order relation} is a relation in the usual sense. 
A \textit{third order relation} of type $\tau =
(r_1,\ldots, r_{w})$ is a set of tuples of (second order) relations of arities $r_1,\ldots, r_{w}$, respectively.
For $i \geq 4$, an \textit{$i$-th order relation} of type $\tau =
(\rho_1,\ldots, \rho_{w})$ is a set of tuples of $(i-1)$-th order relations, of types $\rho_1,\ldots, \rho_{w}$, respectively.
A more formal definition, but also more cumbersome, of the type of higher order relations in this context can be found, e.g., in~\cite{[Fer08]}.
We use uppercase calligraphic letters $\mathcal{X}$, $\mathcal{Y}$, $\mathcal{Z}$, \ldots to denote $i$-th order variables of order $i \geq 3$, uppercase letters $X, Y, Z, \ldots$ to denote second order variables, and lower case letters $x, y, z, \ldots$ to denote first order variables. With $\mathcal{X}^{i,\tau}$ we denote an $i$-th order variable of type $\tau$. For a third order variable $\mathcal{X}^{i,\tau}$, we tend to omit the superindices. We sometimes use $X^r$ to denote that $X$ is a second order variable or arity $r$. 
For any $i \geq 3$, we define the notion of \emph{satisfaction} in HO$^{i}$ as follows: $\mathbf{A},\textit{val} \models \exists \mathcal{X}^{i, (\rho_1,\ldots, \rho_{w})}(\varphi(\mathcal{X}))$, where $\mathcal{X}$ is an $i$-th order relation variable and $\varphi$ is a well-formed formula, if and only if there is at least one  $i$-th order relation $\mathcal{R}$ of type $\tau =
(\rho_1,\ldots, \rho_{w})$  in $A$, such that $\mathbf{A},\mathit{val}^\prime \models \varphi(\mathcal{X})$, where  $\mathit{val}$ and $\mathit{val}^\prime$ are $\mathcal{R}$-equivalent valuations on $\mathbf{A}$, and
$\mathit{val}^\prime(\mathcal{X})=\mathcal{R}$.

Since we use graphs throughout the paper, let us recall in this context that an \emph{undirected graph} is a finite relational structure ${\bf G}$ of vocabulary $\sigma= \{E\}$, satisfying $\varphi_1 \equiv \forall xy (E(x,y) \rightarrow E(y,x))$ and $\varphi_2 \equiv \forall x (\neg E(x,x))$. If we do not require ${\bf G}$ to satisfy both $\varphi_1$, we talk about \emph{directed graph} or \emph{digraph}. We denote by $V$ the domain of the structure $\bf G$, i.e., the set of vertices of the graph ${\bf G}$.

\section{A General Schema of TO Formulae}
\label{sec:schema}

We define next a \textit{general schema} of $\exists$TO formulae which consists of existentially quantifying a third order linear digraph of polynomial length (i.e., a sequence of  structures that represents a computation) by explicitly stating which operations are the ones which can be involved in the construction of a given structure in the sequence, when applied to the previous one. The schema is as follows: 
\begin{align}
\label{eq:generalschema}
\begin{split}
\exists \mathcal{C}^{\bar{s}} \exists \mathcal{O}^{\bar{s} \bar{s}} \; \big(&\text{TotalOrder}({\cal C},{\cal O}) \land \\ & \forall G_1 \forall G_2 \big((\text{First}(G_1)\rightarrow \alpha_{\text{First}}(G_1)) \land (\text{Last}(G_1)\rightarrow \alpha_{\text{Last}}(G_1)) \land \\ & \hspace*{1.45cm} ((\mathcal{C}(G_1) \land \mathcal{C}(G2) \land \text{Pred}(G1,G2)) \rightarrow  \varphi (G1,G2))\big)\big),
\end{split}
\end{align}
where
\vspace{-0.2cm}
\begin{itemize}
\item the relational structures in $\mathcal{C}$ have type $\bar{s} = (i_1, \ldots, i_s)$ with $i_j \ge 1$ for $j = 1,\ldots,s$.
\item $\text{TotalOrder}({\cal C},{\cal O})$, $\text{First}(G_1)$, $\text{Last}(G_1)$ and $\text{Pred}(G1,G2)$ denote fixed SO formulae which express that $\mathcal{O}$ is a total order over $\mathcal{C}$, $G_1$ is the first relational structure in $\mathcal{O}$, $G_1$ is the last relational structure in $\mathcal{O}$ and $G_1$ is the immediate predecessor of $G_2$ in $\mathcal{O}$,  respectively. 
\item $\alpha_{\text{First}}(G_1)$ and $\alpha_{\text{Last}}(G_1)$ denote arbitrary SO formulae which define, respectively, the properties that the first and last structure in $\mathcal{O}$ should satisfy.
\item $\varphi(G1,G2)$ denotes an arbitrary SO formula that expresses how we get $G_2$ out of $G_1$, i.e., which operations can be used to obtain $G_2$ from $G_1$.
\end{itemize}

This is a very usual, intuitive, and convenient schema in the expression of natural properties, as confirmed by the examples in \cite{[FRT14]} and by those discussed through this paper. 

\begin{example}\label{ex:hypercube}
Consider the problem of deciding whether a  graph is a \emph{hypercube}, described as Example~$(a)$ in the introduction.
This can be expressed in TO following the schema (\ref{eq:generalschema}) by letting 
\begin{itemize}
\item $\alpha_{\text{First}}$ express that ``the first graph in the order $\mathcal{O}$ is $K_2$'',
\item $\alpha_{\text{Last}}$ express that  ``the last graph in the order $\mathcal{O}$ is the input graph'', and 
\item $\varphi$ express that ``$G_2$ can be built from two isomorphic copies of $G_1$ by adding edges between the corresponding vertices''.
\end{itemize}
The formula $\varphi$ can be expressed in the following way: Every graph $G_2$ in the sequence results from finding two total, injective functions $f_1, f_2$ from the previous graph $G_1$, so that $(i)$ $f_1$ and $f_2$ induce in $G_2$ two isomorphic copies of $G_1$, $(ii)$ the images of those functions define a partition in the vertex set of $G_2$, and  $(iii)$ there is an edge in $G_2$ between the images $f_1(x)$ and $f_2(x)$ of every node $x$ in $G_1$ (see also the formulae A4.1--A4.5 in~\cite{[FRT14]}, pp.~5 for details).
\end{example}

\begin{example}\label{ex:fvq}
Consider the \textit{Formula-Value query}, described as Example $(c)$ in the Introduction.
Every propositional Boolean formula $\phi$ can be viewed as a \emph{word model}\footnote{For $u=a_1\ldots a_n \in A^+$, a \emph{word model} for $u$ is a structure of the form $(B, <, (P_a)_{a \in A})$ where $|B| = \mathit{length}(u)$, $<$ is a linear order of $B$, and $P_a$ corresponds to the positions in $u$ carrying an $a$ (see~\cite{[EF99]} among others).} $G_{\phi}$ of vocabulary $\pi = \{\le,P_{(},P_{\,)},P_{\land},P_{\lor},P_{\neg},P_{F},P_{T}\}$.
The Formula-Value query can then be expressed in TO following the general schema~(\ref{eq:generalschema}), by letting
\begin{itemize}
\item $\alpha_{\text{First}}$ express that ``the first word model (formula) in $\mathcal{O}$ is the input formula'',
\item $\alpha_{\text{Last}}$ express that ``the last word model (formula) in $\mathcal{O}$ is the formula $T$'', and 
\item $\varphi$ express that ``$G_{\phi_2}$ is obtained by applying to $G_{\phi_1}$ one of the operations of conjunction, disjunction or negation or elimination of a pair of redundant parentheses, which is ready to be evaluated to exactly one subformula of $G_{\phi_1}$''.
\end{itemize}
\end{example}

Interesting additional examples are a set of very relevant relationships between pairs of undirected graphs $\langle G, H\rangle$ defined as orderings of special sorts, which can be expressed following such a schema, by defining a set of possible operations that can be applied repeatedly to $H$, until a graph which is isomorphic to $G$ is obtained. Furthermore, in all these cases, the length of the sequence is at most linear. Namely, the following properties of such kind can be expressed by $\exists$TO formulae that follow the schema described above: $(a)$ $G \leq_{immersion} H$ (i.e., $G$ being an \textit{immersion} in $H$, see~\cite{[AL03],[DF99],[GKMW11]}),
$(b)$ $G \leq_{top} H$ (i.e., $G$ being \textit{topologically embedded}, or \textit{topologically contained} in $H$,  see~\cite{[AL03],[DF99],[GKMW11]}),  $(c)$ $G \leq_{minor} H$ (i.e., $G$ being a \textit{minor} of $H$, see~\cite{[FG06],[DF99]}), $(d)$ $G \leq_{induced-minor} H$ (i.e., $G$ being an \textit{induced minor} of $H$,  see~\cite{[DF99]}). The operations on graphs that are used to define those orderings are (E) delete an edge, (V) delete a vertex, (C) contract an edge, (T) degree 2 contraction, or subdivision removal, and (L) lift an edge. In particular the set of allowable operations for each of those orderings are $\{E, V, L\}$ for $\leq_{immersion}$, $\{E, V, C\}$ for  $\leq_{minor}$, $\{E, V, T\}$ for $\leq_{top}$, and $\{V, C\}$ for $\leq_{induced-minor}$ (see~\cite{[DF99]}).
Another example of the use of a polynomially bounded sequence of structures is the classical definition of \textit{planarity} in undirected graphs. 
one way of stating the classical definition of Kuratowski, that is due to Wagner (1937) and which makes use of one of the orderings mentioned above is the following: a graph is planar if and only if it contains neither $K_{5}$ nor  $K_{3,3}$ as a minor (\cite{[Bol02]}).

\subsection{A Translation of TO formulae of the General Schema to SO}\label{translationOfGeneralSchema}

We show next that if $\Psi$ is a TO formula of the general schema~(1) and there is a polynomial $p$ such that, in all the valuations that satisfy $\Psi$ the cardinality of the third order relation assigned to $\mathcal{C}$ is bounded by $p$ in the size of the input structure, then we can translate $\Psi$ into an equivalent SO formula $\Psi'$.

To simplify the presentation, we first consider the case of graphs and assume that there must be at least one node in each graph in $\cal C$. Let us denote by $d$ and $t$ the degree of the polynomials that bound the number of graphs which can appear in any valuation of ${\cal C}$ which satisfies $\Psi$ and the size of each graph in $\mathcal{C}$, respectively, both in terms of the size $n$ of the (input) structure in which $\Psi$ is evaluated. 

Our strategy consists on encoding the TO relation $\cal C$ as a pair of second order relations $C$ and $E_C$ of arities $d+t$ and $2(d+t)$, respectively. Notice that every formula of the schema~(1) stipulates that $\cal O$ is a linear order of the graphs in $\cal C$ which represents the stages (or steps) of a computation. Consequently the number of stages needed is bounded by $n^d$. Since in turn each stage has a bound on the number of elements it adds or changes (at most $n^t$), we have to consider a set of $(d+t)$-tuples. That is, each graph has at most $n^t$ nodes and we have to allow for sequences of at most $n^d$ stages, where each stage has at most $n^t$ nodes. Regarding the TO relation $\cal O$, we replace it by a pair of second order relations $ST$ and $E_{ST}$, in this case of arities $d$ and $2d$, respectively.

The encoding into second order relations is completed by a left total relation $R \subseteq ST \times C$, where $C$ is the union of the domains of all the structures in the sequence. Every node in $ST$ represents one stage, and, through the forest $R$ defines a subset of nodes, which is the vertex set of a sub graph (not necessarily connected) of the whole graph $(C,E_C)$. We use $C|_{R(\bar{x})}$, ${E_C}|_{R(\bar{x})}$ to denote the restriction of $C$ and $E_C$, respectively, to $R(\bar{x})$, i.e., $C|_{R(\bar{x})}=\{\bar{y} \mid C(\bar{y}) \land R(\bar{x},\bar{y})\}$ and $E_C|_{R(\bar{x})}=\{(\bar{v},\bar{w}) \in E_C \mid R(\bar{x},\bar{v}) \land R(\bar{x},\bar{w})\}$. The sub graph of $(C,E_C)$ which corresponds to the stage $ST(\bar{x})$ is denoted as $(C|_ {R(\bar{x})},E_C|_{R(\bar{x})})$.

Then the translation to SO of TO formulae of the schema~(1) for the case of (non-empty) arbitrary graphs can be done as follows:

\begin{align}
\begin{split}
\exists C^{d+t} E_{C}^{2(d+t)} ST^{d} E_{ST}^{2d} R^{2d+t}\big(&\text{Linear}(ST,E_{ST}) \land R \subseteq ST \times C \land  \text{LeftTotal}(R) \land\\
& \forall\bar{x}\forall\bar{y}\big((\text{First}(\bar{x}) \rightarrow \hat{\alpha}_{\text{First}}) \land (\text{Last}(\bar{x}) \rightarrow \hat{\alpha}_{\text{Last}}) \land \\
& \hspace*{1.2cm}((ST(\bar{x}) \land ST(\bar{y}) \land \text{Pred}(\bar{x}, \bar{y})) \rightarrow \\
& \hspace*{1.0cm} \hat{\varphi}((C|_{R(\bar{x})},E_C|_{R(\bar{x})}),(C|_{R(\bar{y})},E_C|_{R(\bar{y})})))\big)\big),
\end{split}
\end{align}
\noindent where
\vspace{-0.2cm}

\begin{itemize}
\item $\text{Linear}(ST,E_{ST})$, $\text{First}(\bar{x})$, $\text{Last}(\bar{x})$  and $\text{Pred}(\bar{x}, \bar{y}))$ 
denote SO formulae which express that $(ST,E_{ST})$ is a linear digraph, $\bar{x}$ is the first node in $(ST,E_{ST})$, $x$ is the last node in $(ST,E_{ST})$ and $\bar{x}$ is the immediate predecessor of $\bar{y}$ in $(ST,E_{ST})$, respectively. 
\item $R \subseteq ST \times C$ and $\text{LeftTotal}(R)$ are shorthands for $\forall \bar{x} \bar{y} (R(\bar{x}, \bar{y}) \rightarrow (ST(\bar{x}) \wedge C(\bar{y})))$ and $\forall \bar{x} (ST(\bar{x}) \rightarrow \exists \bar{y} (R(\bar{x}, \bar{y})))$, respectively.  
\item $\hat{\alpha}_{\text{First}}$ and $ \hat{\alpha}_{\text{Last}}$ are SO formulae built from $\alpha_{\text{First}}$ and $\alpha_{\text{Last}}$, respectively,  by modifying them to talk about the graph described by $\bar{x}$ through $ST(\bar{x})$, $E_{ST}$ and $R$.
\item $\hat{\varphi}$  is an SO formula built from $\varphi$ by modifying it to talk about the graphs described by $\bar{x}$ and $\bar{y}$ through $ST(\bar{x})$, $ST(\bar{y})$, $E_{ST}$ and $R$.
\end{itemize}

\begin{example}
Take the TO formula described in Example~\ref{ex:fvq} for expressing the Formula-Value query. In this case, $t = 1$, since the size of the input formulae is equal to the size $n$ of the word model presenting it, and the whole evaluation process takes up to $n$ steps.  Its translation to SO using the strategy described in this section then results in a SO formula of the schema~(2) where $(C, E_C)$ encodes a graph whose nodes are binary tuples and whose edges are quadruples, and $(ST, E_{ST})$ encodes a linear digraph of length at most $n$. In turn, $R$ encodes a ternary relation such that $(x, y, z) \in R$ if and only if $(y, z) \in C|_{R(\bar{x})}$. Figure~\ref{fig2} depicts one of the valuations for the key SO variables which satisfies the resulting SO query when it is evaluated over a word model with domain $\{1, 2, \ldots, 8\}$ that encodes the formula $(T \wedge (\neg F))$. 
\begin{figure}[h]
\vspace*{0.7cm}

\centering
\includegraphics[trim=0cm 6.5cm 3cm 6.5cm, width=1.1\textwidth]{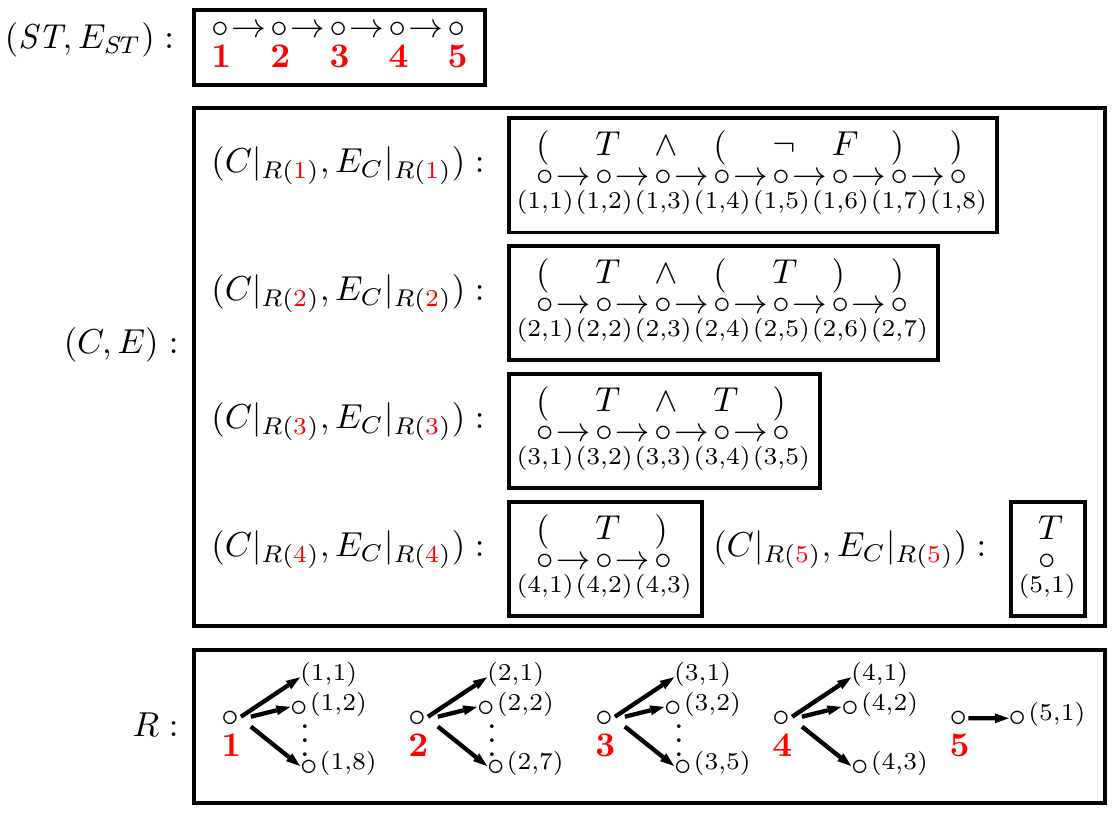}
\vspace{-7.7cm}
\caption{\label{fig2}}
\end{figure}
\vspace*{1.2cm}
\end{example}
For the case of relations of arbitrary arity, say $S$ of arity $r \geq 1$, we simply need to consider $E_C$ as an $r$-ary relation (denoted $E^S_C$). Thus $E_{C}^S|_{R(\bar{x})} = \{(\bar{v_1},\ldots,\bar{v_r}) \in E^S_{C} \colon R(\bar{x},\bar{v_1}) \land \ldots \land R(\bar{x},\bar{v_r})\}$. If we have a tuple of relations, say $l \geq 1$ relations of arities $r_1,\ldots,r_l \geq 1$, respectively, then we have to consider similarly $E_{C_1}^{S_1}$,\ldots,$E_{C_l}^{S_l}$. 

Thus, we get the following (recall that given a relational structure $\bf I$, with $I^{\bar{s}}$ we denote the set of all TO relations (or relational structures) of type (signature) $\bar{s}$ in the set $I = \mathit{dom}(\bf I)$).

\begin{theorem}\label{th1}
Let $\Psi \equiv \exists \mathcal{C}^{\bar{s}} \mathcal{O}^{\bar{s}\bar{s}} \psi(\mathcal{C},\mathcal{O})$ be a TO formula of some relational vocabulary $\sigma$, of the form~(\ref{eq:generalschema}).
There is a translation of $\Psi$ to an equivalent SO formula if the following holds:
\begin{enumerate}[label=\emph{\alph*}.]
\item The subformulae $\alpha_{\text{First}}$, $\alpha_{\text{Last}}$ and $\varphi$ are SO formulae.
 \item There is a positive integer $d$ such that, for every $\sigma$-structure $\bf I$, every TO relation $\mathcal{R}$ in $I^{\bar{s}}$, and every valuation $v$ with $v(\mathcal{C}) = \mathcal{R}$, if  $\mathbf{I}, v \models  \exists \mathcal{O}^{\bar{s}\bar{s}} \psi(\mathcal{C})$, then $|\mathcal{R}| \le |\mathit{dom}(\mathbf{I})|^{d}$.
\end{enumerate}
\end{theorem}

\begin{remark}
If we restrict $\alpha_{\text{First}}$, $\alpha_{\text{Last}}$ and $\varphi$ to $\exists$SO formulae, then Theorem~\ref{th1} can be seen as a direct consequence of Fagin's famous theorem~\cite{[Fag74]} which states that $\exists$SO captures NP. Note that every property definable by a TO formula of the form~(\ref{eq:generalschema}) such that $\alpha_{\text{First}}$, $\alpha_{\text{Last}}$ and $\varphi$ are $\exists$SO formulae and property~(b) in Theorem~\ref{th1} holds, can be checked in NP exactly as it happens for every property definable in SO (it suffices to additionally guess a polynomial-sized valuation for the two existentially quantified TO variable). Then, by Fagin's theorem, we get that every property definable by such kind of TO formulae can also be defined in $\exists$SO.  Nevertheless, the constructive approach that we follow in this paper has the advantage of providing an actual translation to SO which is clear and intuitive, as well as new insight into the problem, in particular if we look at it from the perspective of database query languages.  
\end{remark}

\section{The Fragment TO$^P$ of Third Order Logic}
\label{sec:top}
Now we define a restriction of TO, denoted $\text{TO}^P$, standing for  \textit{polynomial third order logic}. In $\text{TO}^P$ the cardinality of the TO relations which can be assigned by a valuation to a TO variable will be bounded by the degree of a polynomial that depends on the quantifier.
 In the alphabet of $\text{TO}^P$, for every positive integer $d$ we have a third order quantifier $\exists^{P,d}$
and for every third order type $\bar{r}$, we have countably many third order variable symbols  $\mathcal{X}^{d,\bar{r}}$. Here, we will usually avoid the superindex $d$ to simplify the notation.
A \emph{valuation} in a structure $\mathbf{A}$ in this setting assigns to each $\text{TO}^P$  variable $\mathcal{X}^{d,\bar{r}}$ a TO relation $\mathcal{R}$ in $A^{\bar{r}}$, such that $|\mathcal{R}|\le |\mathit{dom}(\mathbf{A})|^d$.
As usual in Finite Model Theory, given that we study logics as a means to express queries to relational structures (which, unless they are Boolean, they define an SO relation in each structure of the corresponding signature) we do not allow free SO or TO variables in $\text{TO}^P$.
The TO$^P$ quantifier $\exists^{P,d}$ has the following semantics:
let $\mathbf{A}$ be a structure; then $\mathbf{A} \models \exists^{P,d} \mathcal{X}^{d,\bar{r}} \varphi(\mathcal{X})$ if and only if there is TO relation $\mathcal{R}^{\bar{r}}$ of type $\bar{r}$, such that $\mathbf{A} \models \varphi(\mathcal{X})[\mathcal{R}]$ and $|\mathcal{R}|\le |\mathit{dom}(\mathbf{A})|^d$.
Therefore we have
\begin{equation}
\label{eq:formula}
\mathbf{A},v \models \exists^{P,d} \mathcal{X}^{d,(r_1,\ldots,r_s)}(\varphi(\mathcal{X}))[\mathcal{R}^{(r_1,\ldots,r_s)}] \; \text{iff} \;\mathbf{A} \models \varphi[\mathcal{R}/\mathcal{X}] \; \text{and} \; |\mathcal{R}| \le \mathit{dom}(\mathbf{A})^d,
\end{equation}
where $\mathbf{A}$ is a structure, $d \geq 1$ is the degree of the polynomial, $\varphi$ is a $\text{TO}^P$  formula of the same signature as $\mathbf{A}$, $\mathcal{X}^{d,(r_1,\ldots,r_s)}$ is a free $\text{TO}^P$ variable in $\varphi$, and $v$ is a valuation which assigns the TO relation $\mathcal{R}^{(r_1,\ldots,r_s)}$, of the same type, to the variable $\mathcal{X}$, i.e., $v(\mathcal{X}) = \mathcal{R}$.

\subsection{A Translation of TO$^p$  formulae to SO}\label{translationOfTOp}

We show next that the above formula (\eqref{eq:formula}), and in general every $\text{TO}^P$ formula, can be translated to an equivalent SO formula.

Note that in classical third order logic, a TO relation $\mathcal{R}^{(r_1,\ldots,r_s)}$ in a structure $\mathbf{A}$ satisfies
$\mathcal{R} \subseteq \mathcal{P}(\mathit{dom}(\mathbf{A})^{r_1}) \times \ldots \times \mathcal{P}(\mathit{dom}(\mathbf{A})^{r_s})$.
Hence, $|\mathcal{R}|\leq 2^{|\mathit{dom}(\mathbf{A})|^{O(1)}}$.
%

\begin{figure}[h!]
\centering
\includegraphics[trim=2.5cm 5.5cm 2.5cm 8.0cm, width=1\textwidth]{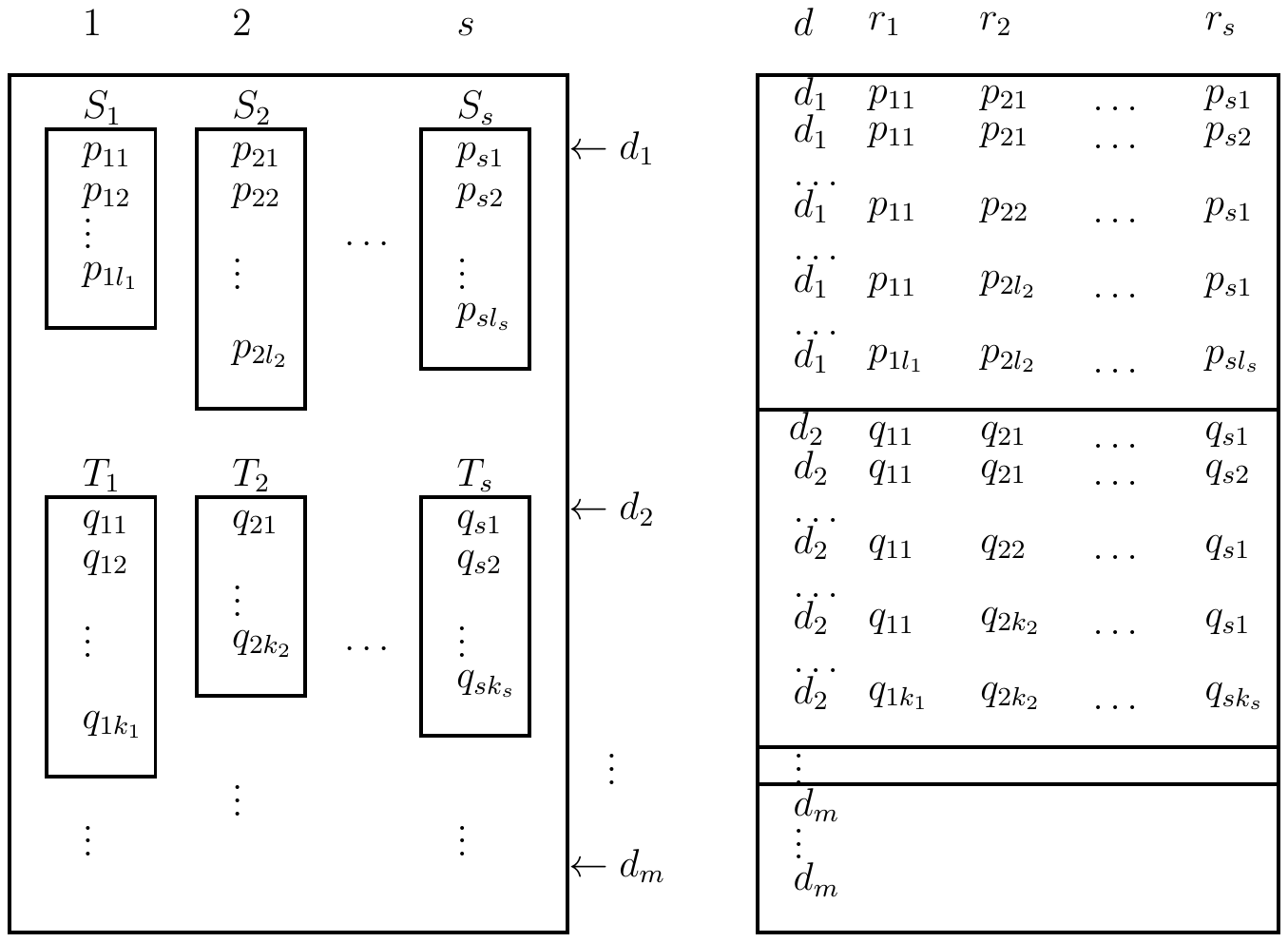}
\squeezeup
\caption{Encoding of the TO relation $\mathcal{R}^{(r_1,\ldots,r_s)}$ to the SO relation $R^{d+r_1+\ldots+r_s}_{\mathcal{R}}$.}
\label{fig:encodTOp}
\end{figure}%

To encode the polynomially bounded TO relations that can be assigned to a TO relation variable in $\text{TO}^P$ we use SO relations as follows. Let
$\mathcal{R}^{(r_1,\ldots,r_s)}$ be a TO relation of type $(r_1,\ldots,r_s)$ as above, and let $d \geq 1$ be the degree of the bounding polynomial such that  $|\mathcal{R}|\le |\mathit{dom}(\mathbf{A})|^d$, for some structure $\mathbf{A}$.
 We use an SO relation $R_{\mathcal{R}}^{d+r_1+\ldots+r_s}$, of arity $(d+r_1+\ldots+r_s)$ to encode $\mathcal{R}^{(r_1,\ldots,r_s)}$, where we use $d$-tuples from $\mathit{dom}(\mathbf{A})^d$ as identifiers of tuples of SO relations in $\mathcal{R}$, so that whenever a tuple $(a_1,\ldots,a_d,a_{d+1},\ldots,a_{d+r_1},\ldots,$\\
$a_{d+r_1+\ldots+r_{s-1}+1},\ldots,a_{d+r_1+\ldots+r_s}) \in R_{\mathcal{R}}$ it means that there is a tuple of SO relations in $\mathcal{R}$ identified by the sub-tuple $(a_1,\ldots,a_d)$, which consists of $s$ SO relations $S_1^{r_1},\ldots,S_s^{r_s}$, of arities  $r_1,\ldots,r_s$, respectively, such that $(a_{d+1},\ldots,a_{d+r_1}) \in S_1,\ldots,(a_{d+r_1+\ldots+r_{s-1}+1},\ldots,
a_{d+r_1+\ldots+r_s}) \in S_s$. This encoding is depicted in Figure~\ref{fig:encodTOp}, where we used the following notation: $(S_1^{r_1},S_2^{r_2},$\\
$\ldots,S_s^{r_s})$, $(T_1^{r_1},T_2^{r_2},\ldots,T_s^{r_s})$ are SO relations in $\mathcal{R}$ of cardinality $l_i=|S_i^{r_i}|$, respectively $k_i=|T_i^{r_i}|$, for $1\le i \le s$. The SO relations $S_i^{r_i}$, $T_i^{r_i}$ contain in turn $r_i$-tuples $p_{i j}$ with $1\le i \le s$ and $1\le j\le l_i$ and respectively $r_i$-tuples $q_{i j}$, with $1\le i \le s$ and $1\le j\le k_i$.

\subsubsection*{Atomic Formulae}

Let $\alpha \equiv  \mathcal{X}^{d,(r_1,\ldots,r_s)}(X_1^{r_1},\ldots,X_s^{r_s})$, with $d, s \geq 1$, $r_1,\ldots,r_s \geq 1$, and   where $\mathcal{X}$ is a $\text{TO}^P$ relation variable of type $(r_1,\ldots,r_s)$.

First, for a better understanding of the translation, let's assume that there are \textit{no empty SO relations} in the tuples of SO relations  in the TO relations that can be assigned by a valuation to the $\text{TO}^P$ relation variable $\mathcal{X}$. Then, considering the encoding of polynomially bounded TO relations described above, every $\text{TO}^P$ relation variable
$\mathcal{X}^{d,(r_1,\ldots,r_s)}$ of type $(r_1,\ldots,r_s)$, which has been quantified by a quantifier $\exists^{P,d}$, with  $d \geq 1$, can be encoded in an SO relation variable $X_{\mathcal{X},{ne}}^{d+r_1+\ldots+r_s}$ of arity $(d+r_1+\ldots+r_s)$ (with the sub-index \textit{ne} in $X_{\mathcal{X},{ne}}$ we denote the restriction assumed above). Accordingly, \textit{in this specific case} the SO formula $\hat{\alpha}_{ne}$ equivalent to $\alpha$ is as follows:\\[0.2cm]
$\hat{\alpha}_{ne} \equiv\exists z_1 \ldots z_{d} \ldots z_{d+r_1+\ldots+r_s}\big(``X_1 \neq \emptyset \land \ldots \land X_s \neq \emptyset\text{''} \land X_{\mathcal{X},{ne}}( z_1,\ldots,z_d,\ldots,$\\
$z_{d+r_1+\ldots+r_s}) \land\forall v_1 \ldots v_{d+r_1+\ldots+r_s}([v_1=z_1 \land \ldots \land v_d =z_d \land X_{\mathcal{X},ne}(v_1,\ldots,v_d,\ldots,$ \\
$v_{d+r_1+\ldots+r_s})] \rightarrow [X_1(v_{d+1},\ldots,v_{d+r_1}) \land \ldots \land  X_s(v_{d+r_1+\ldots+r_{s-1}+1},\ldots,\\
v_{d+r_1+\ldots+r_s})]) \land$ $\forall v_{d+1} \ldots v_{d+r_1+\ldots+r_s}([X_1(v_{d+1},\ldots,v_{d+r_1}) \land \ldots \land  \\
X_s(v_{d+r_1+\ldots+r_{s-1}+1}, \ldots, v_{d+r_1+\ldots+r_s})]$  $\rightarrow \exists v_1 \ldots v_d (v_1=z_1 \land  \ldots \land v_d =z_d \land \\
X_{\mathcal{X},ne}(v_1,\ldots,v_d,\ldots, v_{d+r_1+\ldots+r_s})))\big).$

\vspace*{0.2cm}

However, in the general case we \textit{do have to} consider the cases where either some, all or none of the SO relations that form a given tuple in a TO relation, are empty. Then, instead of encoding  a TO relation $\mathcal{R}^{(r_1,\ldots,r_s)}$ with a single SO relation $R_{\mathcal{R}}^{d+r_1+\ldots+r_s}$, we use several SO relations to encode it. In fact we use exactly $2^{s}$ SO relations, since there are $2^{s}$ possible patterns of empty and non empty relations in a tuple of $s$ SO relations. We denote by $\omega = (i_1,\ldots,i_{|\omega|})$ such a pattern of \textit{empty} relations, with
$1\le i_1 < i_2 <\ldots<i_{|\omega|}\le s$ being the  indices of the components in an $s$-tuple of SO relations that are empty. Correspondingly, we denote by $\bar{\omega} = (j_1,\ldots,j_{|\bar{\omega}|})$ the corresponding pattern of \textit{non empty} relations, with
$1\le j_1 < j_2 < \ldots <j_{|\bar{\omega}|} \le s$ being the  indices of the components in an $s$-tuple of SO relations, that are non empty. By abuse of notation, we will denote as $\{\omega\}$ and $\{\bar{\omega}\}$ the sets of indices in $\omega$ and $\bar{\omega}$, respectively.
Note that the case considered above, where all the components of an $s$-tuple of SO relations are non empty, is one particular value of those patterns, namely $\{\omega\} = \emptyset$ and $\bar{\omega} = (1,\ldots,s)$.

In the formula  $\hat{\alpha}$ for the general case, we need to refer to tuples of varying length, since for each pattern $\omega$, the arity of the SO relation $X_{\mathcal{X},e,\omega}$ which contains the tuples of SO relations in $\mathcal{X}$ (to be precise, in the TO relation assigned to the $\text{TO}^P$ variable $\mathcal{X}$ by a given valuation) whose components with empty relations follow the pattern $\omega$, depends on $\omega$. For that matter we use the following notation (recall that by $\bar{s} \bar{s'}$ we mean the concatenation of the sequences $\bar{s}$ and  $\bar{s'}$): ${\bar{f}}_{\bar{\omega}} = \bar{f}_{j_1} \ldots \bar{f}_{j_{|\bar{\omega}|}}$, where $\bar{f}_{j_1} = (f_{{j_1} 1},\ldots,f_{{j_1} r_{j_1}})$,\ldots, $\bar{f}_{j_{|\bar{\omega}|}} = (f_{{j_{|\bar{\omega}|}} 1},\ldots,f_{{j_{|\bar{\omega}|}} r_{j_{|\bar{\omega}|}}})$. We also use ${\bar{f}^{\prime}}_{\bar{\omega}}$ with the same meaning.

Then, the SO translation of the $\text{TO}^P$ atomic formula $\alpha \equiv  \mathcal{X}^{d,(r_1,\ldots,r_s)}(X_1^{r_1},$\\
$\ldots,X_s^{r_s})$, \textit{in the
general case}, is  the following SO formula $\hat{\alpha}$:
\begin{align*}
&\hat{\alpha} \equiv \bigvee_{\omega \in \Omega} \bigg([``(X_{i_1}=\emptyset \land \ldots \land
X_{i_{|\omega|}}=\emptyset)\text{''} \land ``(X_{j_1}\neq \emptyset \land \ldots \land X_{j_{|\bar{\omega}|}} \neq \emptyset)\text{''}] \land \\
&\big(\exists v_1 \ldots v_d \bar{f}_{\bar{\omega}}\big(X_{\mathcal{X},e,\omega}(v_1,\ldots,v_d,\bar{f}_{\bar{\omega}}) \land \forall u_1 \ldots u_d \bar{f}^{\prime}_{\bar{\omega}} \big[ [ u_1=v_1 \land \ldots \land u_d=v_d \land \\
& X_{\mathcal{X},e,\omega}(u_1,\ldots,u_d,\bar{f}^{\prime}_{\bar{\omega}} )] \rightarrow [\underset{l\in\{j_1,\ldots,j_{|\bar{\omega}|}\}}{\bigwedge}X_l(f_{l 1}^\prime,\ldots,f_{l r_l}^\prime) ] \big] \land \\
& \forall \bar{f}_{\bar{\omega}}^\prime [[\underset{l\in\{j_1,\ldots,j_{|\bar{\omega}|}\}}{\bigwedge}X_l(f_{l 1}^\prime,\ldots,f_{l r_l}^\prime) ] \rightarrow  \exists u_1 \ldots u_d (u_1=v_1 \land \ldots \land u_d=v_d \land \\
& X_{\mathcal{X},e,\omega}(u_1,\ldots,u_d,\bar{f}^{\prime}_{\bar{\omega}} ))] \big) \big)\bigg),
\end{align*}
where $\Omega = \{\omega \mid \omega = (i_1,\ldots,i_{|\omega|})\,;\, 1 \le i_1 < i_2 < \ldots < i_{|\omega|} \le s\,;\,  0\le |\omega| \le s\,;\, \bar{\omega}=(j_1,\ldots,j_{|\bar{\omega}|})\,;\, \{\bar{\omega}\}\cup \{\omega\}=\{1,\ldots,s\}\,;\, \{\bar{\omega}\}\cap\{\omega\}=\emptyset\}$.

\subsubsection*{The Existential Case}

Now, let  $\alpha \equiv\exists^{P,d} \mathcal{X}^{d,(r_1,\ldots,r_s)}(\varphi)$, with $d, s \geq 1$, $r_1,\ldots,r_s \geq 1$, and   where $\mathcal{X}$ is a $\text{TO}^P$ relation variable of type $(r_1,\ldots,r_s)$.
Note that no $d$-tuple can be in more than one of the different SO relations that encode a given polynomially bounded TO relation.
The SO translation of the $\text{TO}^P$ formula $\alpha$ in this case, is  the following SO formula $\hat{\alpha}$:

\begin{align*}
\hat{\alpha} \equiv
&\{\exists X^{d+|\bar{f}_{\bar{\omega}}|}_{\mathcal{X},e,\omega}\}_{\omega \in \Omega} \big(\forall z_1 \ldots z_d
\big[\underset{\underset{\underset{1 \le |\omega| \le s}{1\le i_1 < i_2 \ldots < i_{|\omega|} \le s}}{\omega=(i_1,\ldots,i_{|\omega|})}}{\bigwedge} \forall \bar{f}_{\bar{\omega}} [X_{\mathcal{X},e,\omega}(z_1,\ldots,z_d,\bar{f}_{\bar{\omega}})\rightarrow \\
&(\underset{{\underset{\underset{1 \le |\omega^\prime| \le s; \; {\omega}^\prime \neq \omega}{1\le i^\prime_1 < i^\prime_2 \ldots < i^\prime_{|{\omega}^\prime|} \le s}}{\omega^\prime=(i^\prime_1,\ldots,i^\prime_{|\omega^\prime|})}}}{\bigwedge} \forall \bar{f}_{\bar{\omega}^\prime}^\prime (\neg X_{\mathcal{X},e,{\omega}^\prime}(z_1,\ldots,z_d,\bar{f}_{\bar{\omega}^\prime}^{\prime})))] \big] \big) \land \hat{\varphi},
\end{align*}
where $\Omega = \{\omega \mid \omega = (i_1,\ldots,i_{|\omega|}) \,;\, 1 \le i_1 < i_2 < \ldots < i_{|\omega|} \le s \,;\,  0\le |\omega| \le s \,;\, \bar{\omega}=(j_1,\ldots,j_{|\bar{\omega}|}) \,;\, \{\bar{\omega}\}\cup \{\omega\}=\{1,\ldots,s\} \,;\, \{\bar{\omega}\}\cap\{\omega\}=\emptyset\}$, and $\hat{\varphi}$ is  the SO formula equivalent to the $\text{TO}^P$ formula $\varphi$, obtained by applying inductively the translations described above.

The cases for the translation on logical connectives are trivial.

\subsection{Some Considerations Concerning the Expressive Power of TO$^p$}

The following result is an immediate consequence of the translation presented in the previous section.

\begin{theorem}
\label{thm:topSO}
$\text{TO}^P$ collapses to SO. That is, for every formula in $\text{TO}^P$ there is an equivalent SO formula.
\end{theorem}

The schema of TO formulae introduced in Section~\ref{sec:schema} is a special case of $\text{TO}^P$ formulae, and hence besides the SO translation given in Subsection~\ref{translationOfGeneralSchema}, the TO formulae that follow that schema  have an additional translation, which is the one we used to translate $\text{TO}^P$ formulae to SO in Subsection~\ref{translationOfTOp}. Nevertheless, the translation of Subsection~\ref{translationOfGeneralSchema} yields a more clear and intuitive SO formula, and the maximum arity of the quantified SO relation variables in general seems to be much smaller. For the case of hypercube graphs the maximum arity obtained by the schema translation is $4$, while the SO formulae obtained by the $\text{TO}^P$ translation has maximum arity $8$ ($X_{\mathcal{C}}$  has arity $4$, since the degree is $1$ and the type is $(1,2)$, and hence $X_{\mathcal{O}}$ has arity $8$). And for the case of the Formula-Value query the maximum arity obtained by the schema translation is also $4$, while the SO formulae obtained by the $\text{TO}^P$ translation has maximum arity $22$ ($X_{\mathcal{C}}$  has arity $11$, since the degree is $1$ and the type is $(1,2,1,1,1,1,1,1,1)$, and hence $X_{\mathcal{O}}$ has arity $22$). Note that the maximum arity of a relation symbol in an SO formula \textit{is relevant} for the complexity of its evaluation (see among others~\cite{[HT06]}). Hence, and not surprisingly, it makes sense to study specific schemas of  TO formulae that have equivalent SO formulae, aiming to find more efficient translations than the general strategy used for  $\text{TO}^P$ formulae (which had the purpose of proving equivalence, rather than looking for efficiency in the translation).

In~\cite{[ST06]} we showed that for any $i \geq 3$ the deterministic inflationary fixed-point quantifier ($IFP$) in HO$^{i}$ (i.e., where the variable which is bound by the $IFP$ quantifier is an $(i+1)$-th order variable) is expressible in $\exists$HO$^{i+1}$. Let  $IFP|_P$ denote the restriction of $IFP$ where there is a positive integer $d$ such that in every structure $\mathbf{A}$, the number of stages of the fixed-point is bounded by $|\mathit{dom}(\mathbf{A})|^d$. And let $(SO + IFP)$ denote SO extended with the deterministic inflationary fixed-point quantifier, where the variable which is bound by the $IFP$ quantifier is a third order variable. Note that the addition of such $IFP$ quantifier to SO means that we can express iterations of length exponential in $|\mathit{dom}(\mathbf{A})|$, so that it is strongly conjectured that $(SO + IFP)$ strictly includes SO as to expressive power. However, as a consequence of Theorem~\ref{thm:topSO} above this is not the case with $IFP|_P$. Then, the following corollary is immediate:

\begin{corollary}
$(SO + IFP|_P)$ collapses to SO. That is, for every formula in $(SO + IFP|_P)$  there is an equivalent SO formula. 
\end{corollary}

Finally, let us define $\Sigma{TO^p_n}$ as the restriction of $TO^p$ to prenex formulae of the form $Q_1 V_1 \ldots Q_k V_k (\varphi)$ such that:
\begin{itemize} 
\item $Q_1, \ldots, Q_k \in \{\forall^{P,d}, \exists^{P,d}, \forall, \exists\}$.
\item Each $V_i$ for $1 \leq i \leq k$ is either a second or third order variable (depending on $Q_i$). 
\item $\varphi$ is a first order formula.
\item The prefix $Q_1 V_1 \ldots Q_k V_k$ starts with an existential block of quantifiers and has at most $n$ alternating (between universal and existential) blocks. 
 \end{itemize}

By the well known Fagin-Stockmeyer characterization \cite{[Sto76]} of the polynomial time hierarchy, we know that for every $n \geq 1$ the prenex fragment $\Sigma_n$ of SO captures the level $n$ of the polynomial time hierarchy (denoted $\Sigma^\mathrm{poly}_n$). Using the strategy described in Subsection~\ref{translationOfTOp}, it is not difficult to see that every formula in $\Sigma{TO^p_n}$ can be translated into an equivalent SO formula in $\Sigma_n$. Thus we get the following result:

\begin{theorem}
$\Sigma{TO^P_n}$ captures $\Sigma^\mathrm{poly}_n$.
\end{theorem}


\section{The Fragments $\mathrm{HO^{i,P}}$ of Higher Order Logics}
\label{sec:HOip}

Let $d \ge 1$, and let $\tau =
(r_1,\ldots, r_{w})$  be a
third order relation type.  A \textit{third order relation} $\mathcal{R}$ of type $\tau$ in  a structure $\mathbf{A}$ is \emph{downward polynomially bounded} by $d$ if $|\mathcal{R}|\le |\mathit{dom}(\mathbf{A})|^d$.
Let $i \geq 4$, and let  $\tau =
(\rho_1,\ldots, \rho_{w})$ be an $i$-th order relation type. An \textit{$i$-th order relation} $\mathcal{R}$ of type $\tau$ in  a structure $\mathbf{A}$ is \emph{downward polynomially bounded} by $d$ if $|\mathcal{R}|\le |\mathit{dom}(\mathbf{A})|^d$, and for all $3 \leq j \leq i - 1$, all the $j$-th order relations that form the tuples of $j + 1$-th order relations, are in turn downward polynomially bounded by $d$.

We define inductively a restriction of HO$^{i}$ for every $i \geq 4$. We denote it as HO$^{i, P}$, standing for  \textit{polynomial $i$-th order logic}.
For $i = 4$, HO$^{4,P}$ is  the extension of TO$^{P}$, where the $4$-th order quantifiers restrict the cardinality (i.e., the number of tuples of third order relations of the valuating fourth order relations) to be bounded by a polynomial that depends on the quantifier. Likewise, for $i \geq 5$ we define HO$^{i,P}$ as the extension of HO$^{i - 1,P}$, where the $i$-th order quantifiers restrict the cardinality (i.e., the number of tuples of $(i - 1)$-th order relations  of the valuating $i$-th order relations) to be  bounded by a polynomial that depends on the quantifier.

 In the alphabet of HO$^{i, P}$, for every pair of positive integers $d$, and $j$, with $i \geq j \geq 4$, we have a $j$-th order quantifier $\exists^{j, P,d}$
and for every $j$-th order type $\tau$, we have countably many $j$-th order variable symbols  $\mathcal{X}^{j, d, \tau}$. Here, we will usually avoid the superindices  $d$ and $\tau$ for clarity.
For simplicity we assume that the types of all relations of all orders $3 \leq j \leq i$ in every $i$-th order relation assigned by a valuation to an $i$-th order relation variable, have width $s$, for some $s \geq 1$, and that every such relation is \textit{downward polynomially bounded} by $d \ge 1$.

A \emph{valuation} in a structure $\mathbf{A}$ in this setting assigns to each $i$-th order relation variable $\mathcal{X}^{j, d, \tau}$ an $i$-th order relation $\mathcal{R}$ in $A$, such that $|\mathcal{R}|\le |\mathit{dom}(\mathbf{A})|^d$.
As usual in Finite Model Theory, given that we study logics as a means to express queries to relational structures (which unless they are Boolean, they define a SO relation in each structure of the corresponding signature) we do not allow free SO or $i$-th order relation variables, for any $i \geq 3$,  in HO$^{i, P}$.

For any $3 \leq j \leq i$, the HO$^{i, P}$ quantifier $\exists^{j, P,d}$ has the following semantics:
let $\mathbf{A}$ be a structure, and let $\mathcal{X}^{j, d, \tau}$ be a $j$-th order relation variable; then $\mathbf{A} \models \exists^{j, P,d} \mathcal{X}^{j, d, \tau} \varphi(\mathcal{X})$ if and only if there is a $j$-th order relation $\mathcal{R}$ of type $\tau$, such that $\mathbf{A} \models \varphi(\mathcal{X})[\mathcal{R}]$ and $\mathcal{R}$ is downward polynomially bounded by $d$ in $\mathbf{A}$.


\subsection{Collapse of The Fragment $\mathrm{HO^{i,P}}$ to SO}

We discuss next how to build for every $\mathrm{HO}^{4,P}$ formula $\alpha$ an SO formula which is equivalent to $\alpha$. To that end,  we will have for every fourth order relation variable a set of SO relation variables that represent it. We do so by representing the fourth order relation variable by what in the field of Database Theory is known as a \textit{normalized relational database}.

We use a rather cumbersome notation, mainly for the sub indices in the SO formulae. The aim is to make it very clear that both the names of the different variables needed and the structure of the formulae can be iterated in a straightforward way for \textit{any order} $i \geq 5$, getting thus the corresponding translations for $\mathrm{HO^{i,P}}$ formulae to SO.


Suppose the $\mathrm{HO}^{4,P}$ formula $\alpha$ is  of the form $\mathcal{X}^{4,d,\tau}(\mathcal{Y}_1^3,\ldots,\mathcal{Y}_s^3)$, with $|\tau| = s$, and where all the fourth order relations which valuate $\mathcal{X}$ are assumed to be polynomially bounded with degree $d \ge 1$.
Next, we show that we can build an SO formula which is equivalent to $\alpha$. 

\begin{figure}[!htp]
 \hspace*{-2.2cm}
\includegraphics[scale=0.8]{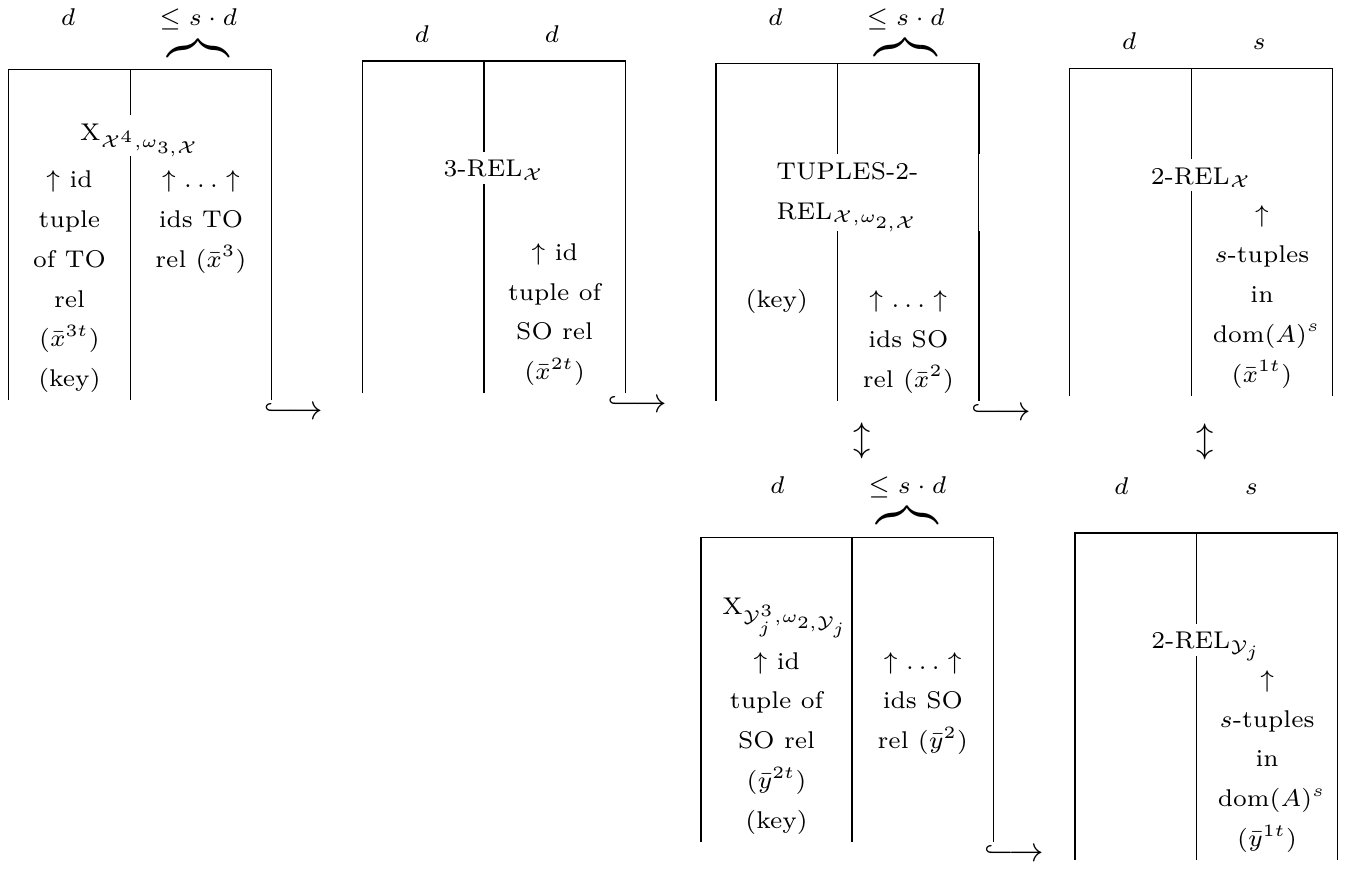}
  \vspace*{-14cm}
  \label{figure:fig1}
  \caption{}
\end{figure}

In Figure~\ref{figure:fig1} we depict the (SO) relation variables that are used to represent the fourth order relation variable $\mathcal{X}^{4,d,\tau}$, and each of the  third order relation variables $\mathcal{Y}_j$ in the tuple $(\mathcal{Y}_1^3,\ldots,\mathcal{Y}_s^3)$.
For $\mathcal{X}^{4,d,\tau}$, we use the following relation variables: 3-REL$_\mathcal{X}$, 2-REL$_\mathcal{X}$, $\mathrm{X}_{{\mathcal{X}^{4},{\omega}_{3,\mathcal{X}}}}$
for each of the different patterns of empty third order relations  ${\omega}_{3,\mathcal{X}}$, and TUPLES-2-REL$_{\mathcal{X},\omega_{2,\mathcal{X}}}$  for each of the different patterns of empty second order relations $\omega_{2,\mathcal{X}}$. In turn, for each relation variable $\mathcal{Y}_j$ in the tuple $(\mathcal{Y}_1^3,\ldots,\mathcal{Y}_s^3)$, we use 2-REL$_{\mathcal{Y}_j}$, and $\mathrm{X}_{{{{\mathcal{Y}^3}_{j_{1_{3,\mathcal{X}}}}}}, {\omega}_{2,{\mathcal{Y}_{{j_{1_{3,\mathcal{X}}}}}}}}$
for each of the different patterns of empty second order relations ${\omega}_{2,{\mathcal{Y}_{j_{1_{3,\mathcal{X}}}}}}$. 

Note that for the individual (first order) variables and tuples we use the following convention. $\bar{x}^{3t}$, $\bar{x}^{3}$, $\bar{x}^{2t}$, $\bar{x}^{2}$ and $\bar{x}^{1t}$ are associated to the encoding of $\mathcal{X}^4$ and denote variables that range, respectively, over identifiers of tuples of TO relations, identifiers of TO relations, identifiers of tuples of SO relations, identifiers of SO relations, and identifiers of tuples of individual elements from the interpreting structure. Likewise, $\bar{y}^{2t}$, $\bar{y}^{2}$ and $\bar{y}^{1t}$ are associated to the encoding of $\mathcal{Y}^3$ and denote variables that range, respectively, over identifiers of tuples of SO relations, identifiers of SO relations, and identifiers of tuples of individual elements from the interpreting structure.
The tuples of variables $\bar{x}^{3t}$, $\bar{x}^{2t}$, $\bar{y}^{2t}$ are of width $d$, while the tuples of variables $\bar{x}^{1t}$ and  $\bar{y}^{1t}$ are of width $s$.
Regarding the patterns of empty relations in tuples of relations, we use the same notation as in $\text{TO}^P$.
  \vspace*{-3cm}

As in the case of $\mathrm{TO}^{P}$, the idea is to represent the fourth order relation variable $\mathcal{X}^{4,d,\tau}$ using $2^s$ (SO) relation variables, one for each pattern of empty TO relations in the tuples of TO relations in the given valuating fourth order relation. In each such relation variable we have the tuples of (non empty) TO relations whose pattern is the one which is used as sub index in the name of the relation variable.
In 3-REL$_\mathcal{X}$ we have the identifiers of the tuples of SO relations which form each TO relation in each of the tuples in $\mathrm{X}_{{\mathcal{X}^{4},{\omega}_{3,\mathcal{X}}}}$.
To represent the tuples of SO relations, we use $2^s$ relation variables, one for each pattern of empty SO relations in the tuples of SO relations in each TO relation in each of the tuples in the different $\mathrm{X}_{{\mathcal{X}^{4},{\omega}_{3,\mathcal{X}}}}$ relation variables.  In each such relation variable we have the tuples of (non empty) SO relations whose pattern is the one which is used as sub index in the name of the relation.
In the different relation variables TUPLES-2-REL$_{\mathcal{X},\omega_{2,\mathcal{X}}}$, we have the identifiers of the SO relations in each tuple of SO relations as above.
Finally, in 2-REL$_\mathcal{X}$ we have all the tuples of individual elements from the interpreting structure which form each SO relation in  the different relation variables TUPLES-2-REL$_{\mathcal{X},\omega_{2,\mathcal{X}}}$.

A detailed explanation of the proposed SO formulae for the nontrivial atomic and existential cases (as well as the actual SO formulae) can be found in Appendices~\ref{sectionA} and~\ref{sectionB}, respectively. It is not difficult to see that the SO formulae used in the translation above for $\mathrm{HO^{4,P}}$ can be iterated, and thus we can build corresponding formulae for any order $i \geq 3$. So we get the following result (as for $\text{TO}^P$, the cases for the translation on logical connectives are trivial).

\begin{theorem}
\label{thm:hopSO}
For all $i \geq 3$, $\mathrm{HO^{i,P}}$ collapses to SO. That is, for every formula in $\mathrm{HO^{i,P}}$ there is an equivalent SO formula which can be built following the translation given above. $\Box$
\end{theorem}

\section{Conclusion}

We think \ref{thm:hopSO} is an interesting result, since beyond the practical applications mentioned in the paper, it means that in the framework of computable queries, where queries define (SO) relations on the input structures, \textit{nesting} in \textit{any arbitrary depth} is \emph{irrelevant} as to expressive power. That is, the only reason why (unrestricted) higher order quantification increases the expressive power of a logic (which essentially means rising the data complexity from non deterministic hyperexponential time of level $(i-2)$ to non deterministic hyperexponential time of level $(i-1)$) is the fact that an $(i+1)$-th order relation can contain an exponential number of tuples of $(i)$-th order relations (see the proofs of the expressive power of HO$^{i}$ in~\cite{[HT06]}).

Note that this fact also appears, among other subjects, in the study of the strict hierarchy induced in the class of primitive recursive functions, by bounding the minimal depth of nesting of the LOOP constructions needed by a LOOP program which can compute a given function in the context of Computability Theory (see~\cite{[DW83]}). The results suggest that, given that the contents of the variable which controls the LOOP iteration is fixed at the beginning of it, LOOP nesting seems to be the only way by which we can increase the running time of a program, on a given input.

\bibliographystyle{plain}
\bibliography{references}

\appendix
\section{The case of Atomic Formulae in $\mathrm{HO}^{4,P}$}\label{sectionA}

In the SO formula below, in the feet descriptions of the big disjunctions and big conjunctions we use simplified expressions with the following meanings:
In the big disjunction of the first line of the formula with ${\omega}_{3,\mathcal{X}}$ we mean the following:
\begin{align*}
&{\omega_{3,\mathcal{X}}=(i_1,\ldots,i_{|\omega_{3,\mathcal{X}}|})},{1\le i_1 < i_2 <\ldots<i_{|\omega_{3,\mathcal{X}}|}\le s},{0\le |\omega_{3,\mathcal{X}}| \le s},\\
&\bar{\omega}_{3,\mathcal{X}}=(j_1,\ldots,j_{|\bar{\omega}_{3,\mathcal{X}}|}),1 \leq j_1 < j_2 <  \ldots < j_{| \bar{\omega}_{3,\mathcal{X}}|} \leq s, \\
&\{\bar{\omega}_{3,\mathcal{X}}\}\cup \{\omega_{3,\mathcal{X}}\}=\{1,\ldots,s\}, \{\bar{\omega}_{3,\mathcal{X}}\}\cap\{\omega_{3,\mathcal{X}}\}=\emptyset.
\end{align*}

In the first big conjunction of the rest of the formula with ${\omega}_{2,{\mathcal{Y}_{j_{1_{3,\mathcal{X}}}}}}$ we mean the following:
\begin{align*}
&{{{\omega}_{2,{\mathcal{Y}_{j_{1_{3,\mathcal{X}}}}}}}=(i_1,\ldots,i_{|{{\omega}_{2,{\mathcal{Y}_{j_{1_{3,\mathcal{X}}}}}}}|})},{1\le i_1 < i_2 <\ldots<i_{|{{\omega}_{2,{\mathcal{Y}_{j_{1_{3,\mathcal{X}}}}}}}|}\le s},\\
&{0\le |{{\omega}_{2,{\mathcal{Y}_{j_{1_{3,\mathcal{X}}}}}}}| \le s},{\bar{\omega}_{2,{\mathcal{Y}_{j_{1_{3,\mathcal{X}}}}}}}=(j_1,\ldots,j_{|\bar{\omega}_{2,{\mathcal{Y}_{j_{1_{3,\mathcal{X}}}}}}|}),\\
&1 \leq j_1 < j_2 <  \ldots < j_{| \bar{\omega}_{2,{\mathcal{Y}_{j_{1_{3,\mathcal{X}}}}}}|} \leq s, \{{\bar{\omega}_{2,{\mathcal{Y}_{j_{1_{3,\mathcal{X}}}}}}}\}\cup \{{{\omega}_{2,{\mathcal{Y}_{j_{1_{3,\mathcal{X}}}}}}}\}=\{1,\ldots,s\},\\
&\{{{\bar{\omega}_{2,{\mathcal{Y}_{j_{1_{3,\mathcal{X}}}}}}}}\}\cap\{{{\omega}_{2,{\mathcal{Y}_{j_{1_{3,\mathcal{X}}}}}}}\}=\emptyset.
\end{align*}
In the second big conjunction of the rest of the formula with ${{\omega}_{2,{\mathcal{Y}_{{j_{|{\bar{\omega}}_{3,\mathcal{X}}|}}}}}}$ we mean the following:
\begin{align*}
&{{{\omega}_{2,{\mathcal{Y}_{{j_{|{{\bar{\omega}}_{3,\mathcal{X}}}|}}}}}}=(i_1,\ldots,i_{|{{\omega}_{2,{\mathcal{Y}_{{j_{|{\bar{\omega}}_{3,\mathcal{X}}|}}}}}}|})},{1\le i_1 < i_2 <\ldots<i_{|{{\omega}_{2,{\mathcal{Y}_{{j_{|{\bar{\omega}}_{3,\mathcal{X}}|}}}}}}|}\le s},\\
&{0\le |{{\omega}_{2,{\mathcal{Y}_{{j_{|{\bar{\omega}}_{3,\mathcal{X}}|}}}}}}| \le s},{\bar{\omega}_{2,{\mathcal{Y}_{{j_{|{\bar{\omega}}_{3,\mathcal{X}}|}}}}}}=(j_1,\ldots,j_{|{\bar{\omega}_{2,{\mathcal{Y}_{{j_{|{\bar{\omega}}_{3,\mathcal{X}}|}}}}}}|}),\\
&1 \leq j_1 < j_2 <  \ldots < j_{| {\bar{\omega}_{2,{\mathcal{Y}_{{j_{|{\bar{\omega}}_{3,\mathcal{X}}|}}}}}}|} \leq s ,\{{\bar{\omega}_{2,{\mathcal{Y}_{{j_{|{\bar{\omega}}_{3,\mathcal{X}}|}}}}}}\}\cup \{{{\omega}_{2,{\mathcal{Y}_{{j_{|{\bar{\omega}}_{3,\mathcal{X}}|}}}}}}\}=\{1,\ldots,s\},\\
&\{{\bar{\omega}_{2,{\mathcal{Y}_{{j_{|{\bar{\omega}}_{3,\mathcal{X}}|}}}}}}\}\cap\{{{\omega}_{2,{\mathcal{Y}_{{j_{|{\bar{\omega}}_{3,\mathcal{X}}|}}}}}}\}=\emptyset.
\end{align*}

We use the same simplification in the feet descriptions of the big disjunctions and big conjunctions in the SO formula for the case of the $\mathrm{HO}^{4,P}$ existential formula.

First we express the formula in a more intuitive way, using natural language, and then we present the actual SO formula with labels in the left side that correspond to the different subformulae in the natural language expression.

\vspace{.3cm}

\noindent \textbf{Natural language expression of the SO formula for Atomic $\mathrm{HO}^{4,P}$ Formulae:}

 \vspace{.3cm}

\noindent
 For \textit{some} pattern of non empty relations for tuples of TO relations ${\omega}_{3,\mathcal{X}}$, the tuple of TO relations $(\mathcal{Y}_1^3,\ldots,\mathcal{Y}_s^3)$ follows that pattern, and there is a tuple of TO relations in the relation $\mathrm{X}_{{\mathcal{X}^{4},{\omega}_{3,\mathcal{X}}}}$, and corresponding  $|{\bar{\omega}_{{3,\mathcal{X}}}}|$ tuples of SO relations in the relation $3\text{-REL}_{\mathcal{X}}$ such that

 \vspace{.3cm}

\noindent \textbf{\underline{I)}}   for \textit{all}    the patterns of non empty relations for tuples of SO relations ${\omega}_{2,{\mathcal{Y}_{j_{1_{3,\mathcal{X}}}}}}$ for the \textit{\textbf{first}} non empty TO relation  ${\mathcal{Y}_{j_{1}}}$  in the tuple  $(\mathcal{Y}_1^3,\ldots,\mathcal{Y}_s^3)$ according to the specific pattern  ${\omega}_{3,\mathcal{X}}$, it so happens that


 \begin{enumerate}

\item
     \textbf{\underline{I.1)}} whenever there is a tuple of SO relations in the relation $\mathrm{X}_{{{{\mathcal{Y}^3}_{j_{1_{3,\mathcal{X}}}}}}, {\omega}_{2,{\mathcal{Y}_{{j_{1_{3,\mathcal{X}}}}}}}}$ (i.e., the SO relation that encodes the relation ${\mathcal{Y}_{j_{1}}}$  mentioned above for the specific pattern ${\omega}_{2,{\mathcal{Y}_{{j_{1_{3,\mathcal{X}}}}}}}$), \textit{then} there is also a tuple of SO relations in the relation $3\text{-REL}_{\mathcal{X}}$ for the \textit{\textbf{first}} non empty TO relation in the tuple of TO relations in the relation $\mathrm{X}_{{\mathcal{X}^{4},{\omega}_{3,\mathcal{X}}}}$ according to the specific pattern  ${\omega}_{3,\mathcal{X}}$, with a corresponding tuple in the relation $\text{TUPLES-2-REL}_{\mathcal{X}}$, such that

 \begin{enumerate}

\item

\begin{enumerate}

\item
        whenever there is a tuple in the relation $2\text{-REL}_{\mathcal{Y}_{j_{1_{3,\mathcal{X}}}}}$ for the \textit{\textbf{first}} non empty SO relation in the current tuple of SO relations in $\mathrm{X}_{{{{\mathcal{Y}^3}_{j_{1_{3,\mathcal{X}}}}}}, {\omega}_{2,{\mathcal{Y}_{{j_{1_{3,\mathcal{X}}}}}}}}$ according to the specific pattern  ${\omega}_{2,{\mathcal{Y}_{j_{1_{3,\mathcal{X}}}}}}$, then there is also a corresponding tuple in the relation $2\text{-REL}_{\mathcal{X}}$ for the \textit{\textbf{first}} non empty SO relation in the current tuple of SO relations in $\text{TUPLES-2-REL}_{\mathcal{X}}$ according to the specific pattern  ${\omega}_{2,{\mathcal{Y}_{j_{1_{3,\mathcal{X}}}}}}$, such that the tuples of elements in the relations $2\text{-REL}_{\mathcal{Y}_{j_{1_{3,\mathcal{X}}}}}$ and $2\text{-REL}_{\mathcal{X}}$ are the same,

\vspace{.3cm}

       and  \ldots   and

\vspace{.3cm}

 \item   whenever there is a tuple in the relation $2\text{-REL}_{\mathcal{Y}_{j_{1_{3,\mathcal{X}}}}}$ for the \textit{\textbf{last}} non empty SO relation in the current tuple of SO relations in $\mathrm{X}_{{{{\mathcal{Y}^3}_{j_{1_{3,\mathcal{X}}}}}}, {\omega}_{2,{\mathcal{Y}_{{j_{1_{3,\mathcal{X}}}}}}}}$ according to the specific pattern  ${\omega}_{2,{\mathcal{Y}_{j_{1_{3,\mathcal{X}}}}}}$, then there is also a corresponding tuple in the relation $2\text{-REL}_{\mathcal{X}}$ for the \textit{\textbf{last}} non empty SO relation in the current tuple of SO relations in $\text{TUPLES-2-REL}_{\mathcal{X}}$ according to the specific pattern  ${\omega}_{2,{\mathcal{Y}_{j_{1_{3,\mathcal{X}}}}}}$, such that the tuples of elements in the relations $2\text{-REL}_{\mathcal{Y}_{j_{1_{3,\mathcal{X}}}}}$ and $2\text{-REL}_{\mathcal{X}}$ are the same,

\end{enumerate}

\vspace{.3cm}

\item                                   and  $\langle$\textit{viceversa}, i.e.,$\rangle$

\vspace{.3cm}

\begin{enumerate}

\item

 whenever there is a tuple in the relation $2\text{-REL}_{\mathcal{X}}$ for the \textit{\textbf{first}} non empty SO relation in the current tuple of SO relations in $\text{TUPLES-2-REL}_{\mathcal{X}}$  according to the specific pattern  ${\omega}_{2,{\mathcal{Y}_{j_{1_{3,\mathcal{X}}}}}}$, then there is also a corresponding tuple in the relation $2\text{-REL}_{\mathcal{Y}_{j_{1_{3,\mathcal{X}}}}}$, such that the tuples of elements in the relations $2\text{-REL}_{\mathcal{Y}_{j_{1_{3,\mathcal{X}}}}}$ and $2\text{-REL}_{\mathcal{X}}$ are the same,

 \vspace{.3cm}

       and  \ldots   and

\vspace{.3cm}

\item
    whenever there is a tuple in the relation $2\text{-REL}_{\mathcal{X}}$  for the \textit{\textbf{last}} non empty SO relation in  the current tuple of SO relations in $\text{TUPLES-2-REL}_{\mathcal{X}}$ according to the specific pattern  ${\omega}_{2,{\mathcal{Y}_{j_{1_{3,\mathcal{X}}}}}}$,  then there is also a corresponding tuple in the relation $2\text{-REL}_{\mathcal{Y}_{j_{1_{3,\mathcal{X}}}}}$, such that the tuples of elements in the relations $2\text{-REL}_{\mathcal{Y}_{j_{1_{3,\mathcal{X}}}}}$ and $2\text{-REL}_{\mathcal{X}}$ are the same,

\end{enumerate}

\end{enumerate}

\vspace{.3cm}


and       $\langle$\textit{viceversa w.r.t. \textbf{(I.1)}}, i.e.,$\rangle$

\vspace{.3cm}

\item
   \textbf{\underline{I.2)}} whenever there is a tuple of SO relations in the relation $3\text{-REL}_{\mathcal{X}}$ for the \textit{\textbf{first}} non empty TO relation in the tuple of TO relations in the relation $\mathrm{X}_{{\mathcal{X}^{4},{\omega}_{3,\mathcal{X}}}}$ according to the specific pattern  ${\omega}_{3,\mathcal{X}}$,  with a corresponding tuple in the relation $\text{TUPLES-2-REL}_{\mathcal{X}}$ \textit{then} there is also a tuple of SO relations in the relation $\mathrm{X}_{{{{\mathcal{Y}^3}_{j_{1_{3,\mathcal{X}}}}}}, {\omega}_{2,{\mathcal{Y}_{{j_{1_{3,\mathcal{X}}}}}}}}$,
    such that

\vspace{.3cm}

$\langle$\textit{the following sub-formula in (a)i, (a)ii, (b)i and (b)ii is the exact repetition of the sub-formula in (a)i, (a)ii, (b)i and (b)ii in \textbf{(I.1)} above}, i.e.,$\rangle$

\vspace{.3cm}

\begin{enumerate}

\item

\begin{enumerate}

\item
        whenever there is a tuple in the relation $2\text{-REL}_{\mathcal{Y}_{j_{1_{3,\mathcal{X}}}}}$ for the \textit{\textbf{first}} non empty SO relation in the current tuple of SO relations in $\mathrm{X}_{{{{\mathcal{Y}^3}_{j_{1_{3,\mathcal{X}}}}}}, {\omega}_{2,{\mathcal{Y}_{{j_{1_{3,\mathcal{X}}}}}}}}$ according to the specific pattern  ${\omega}_{2,{\mathcal{Y}_{j_{1_{3,\mathcal{X}}}}}}$, then there is also a corresponding tuple in the relation $2\text{-REL}_{\mathcal{X}}$ for the \textit{\textbf{first}} non empty SO relation in the current tuple of SO relations in $\text{TUPLES-2-REL}_{\mathcal{X}}$ according to the specific pattern  ${\omega}_{2,{\mathcal{Y}_{j_{1_{3,\mathcal{X}}}}}}$, such that the tuples of elements in the relations $2\text{-REL}_{\mathcal{Y}_{j_{1_{3,\mathcal{X}}}}}$ and $2\text{-REL}_{\mathcal{X}}$ are the same,

\vspace{.3cm}

       and  \ldots   and

\vspace{.3cm}

 \item   whenever there is a tuple in the relation $2\text{-REL}_{\mathcal{Y}_{j_{1_{3,\mathcal{X}}}}}$ for the \textit{\textbf{last}} non empty SO relation in the current tuple of SO relations in $\mathrm{X}_{{{{\mathcal{Y}^3}_{j_{1_{3,\mathcal{X}}}}}}, {\omega}_{2,{\mathcal{Y}_{{j_{1_{3,\mathcal{X}}}}}}}}$ according to the specific pattern  ${\omega}_{2,{\mathcal{Y}_{j_{1_{3,\mathcal{X}}}}}}$, then there is also a corresponding tuple in the relation $2\text{-REL}_{\mathcal{X}}$ for the \textit{\textbf{last}} non empty SO relation in the current tuple of SO relations in $\text{TUPLES-2-REL}_{\mathcal{X}}$ according to the specific pattern  ${\omega}_{2,{\mathcal{Y}_{j_{1_{3,\mathcal{X}}}}}}$, such that the tuples of elements in the relations $2\text{-REL}_{\mathcal{Y}_{j_{1_{3,\mathcal{X}}}}}$ and $2\text{-REL}_{\mathcal{X}}$ are the same,

\end{enumerate}

\vspace{.3cm}

\item                                   and  $\langle$\textit{viceversa}, i.e.,$\rangle$

\vspace{.3cm}

\begin{enumerate}

\item

 whenever there is a tuple in the relation $2\text{-REL}_{\mathcal{X}}$ for the \textit{\textbf{first}} non empty SO relation in the current tuple of SO relations in $\text{TUPLES-2-REL}_{\mathcal{X}}$  according to the specific pattern  ${\omega}_{2,{\mathcal{Y}_{j_{1_{3,\mathcal{X}}}}}}$, then there is also a corresponding tuple in the relation $2\text{-REL}_{\mathcal{Y}_{j_{1_{3,\mathcal{X}}}}}$, such that the tuples of elements in the relations $2\text{-REL}_{\mathcal{Y}_{j_{1_{3,\mathcal{X}}}}}$ and $2\text{-REL}_{\mathcal{X}}$ are the same,

 \vspace{.3cm}

       and  \ldots   and

\vspace{.3cm}

\item
    whenever there is a tuple in the relation $2\text{-REL}_{\mathcal{X}}$  for the \textit{\textbf{last}} non empty SO relation in  the current tuple of SO relations in $\text{TUPLES-2-REL}_{\mathcal{X}}$ according to the specific pattern  ${\omega}_{2,{\mathcal{Y}_{j_{1_{3,\mathcal{X}}}}}}$,  then there is also a corresponding tuple in the relation $2\text{-REL}_{\mathcal{Y}_{j_{1_{3,\mathcal{X}}}}}$, such that the tuples of elements in the relations $2\text{-REL}_{\mathcal{Y}_{j_{1_{3,\mathcal{X}}}}}$ and $2\text{-REL}_{\mathcal{X}}$ are the same,

\end{enumerate}

\end{enumerate}

\vspace{.3cm}

 and  \ldots   and

\vspace{.3cm}

\end{enumerate}

\vspace{1cm}

\noindent \textbf{\underline{II)}}  $\langle$\textit{the following subformula is an \textbf{exact repetition} of the sub-formula \textbf{(I)}, \textbf{except} that the subindex ${{j_{1_{3,\mathcal{X}}}}}$ must be replaced by
 ${j_{|{\bar{\omega}}_{3,\mathcal{X}}|}}$ in all its occurrences,} i.e.,$\rangle$

\vspace{.3cm}

 for \textit{all}    the patterns of non empty relations for tuples of SO relations ${\omega}_{2,{\mathcal{Y}_{{j_{|{\bar{\omega}}_{3,\mathcal{X}}|}}}}}$ for the \textit{\textbf{last}} non empty relation  ${\mathcal{Y}_{{j_{|{\bar{\omega}}_{3,\mathcal{X}}|}}}}$  in the tuple  $(\mathcal{Y}_1^3,\ldots,\mathcal{Y}_s^3)$ according to the specific pattern  ${\omega}_{3,\mathcal{X}}$, it so happens that...

\vspace{.3cm}

$\langle$\textit{then, correspondingly, t\textbf{he following labels would follow:} 1.II.1, 1(a)i, 1(a)ii, 1(b)i, 1(b)ii, 2.II.2, 2(a)i, 2(a)ii, 2(b)i, 2(b)ii}$\rangle$.

 \vspace{.3cm}




































\vspace{.3cm}

\noindent \textbf{SO formula for Atomic $\mathrm{HO}^{4,P}$ Formulae:}

 \vspace{.3cm}

\[\bigvee_{{\omega}_{3,\mathcal{X}}}  \exists{\bar{x}}^{3t} {\bar{x}^{3}_{j_{1_{3,\mathcal{X}}}}} \ldots  {\bar{x}}^{3}_{j_{|{\bar{\omega}_{{3,\mathcal{X}}}}|}} \bar{x}^{2 t}_{j_{1_{3,\mathcal{X}}}} \ldots \bar{x}^{2t}_{j_{|{\bar{\omega}_{{3,\mathcal{X}}}}|}} \bigg((\text{``} {\mathcal{Y}^{3}_{i_{1_{3,\mathcal{X}}}} = \emptyset } \text{''} \land \ldots \land \text{``} {\mathcal{Y}^{3}_{i_{|{{\omega}_{{3,\mathcal{X}}}}|}} = \emptyset} \text{''}) \land \]
\[(\text{``} {\mathcal{Y}^{3}_{j_{1_{3,\mathcal{X}}}} \neq \emptyset } \text{''} \land \ldots \land \text{``} {\mathcal{Y}^{3}_{j_{|{\bar{\omega}_{{3,\mathcal{X}}}}|}} \neq \emptyset} \text{''}) \land \mathrm{X}_{{\mathcal{X}^{4},{\omega}_{3,\mathcal{X}}}}(\bar{x}^{3t},{\bar{x}^{3}_{j_{1_{3,\mathcal{X}}}}}, \ldots,  {\bar{x}}^{3}_{j_{|{\bar{\omega}_{{3,\mathcal{X}}}}|}}) \land \]
\[3\text{-REL}_{\mathcal{X}}({\bar{x}^{3}_{j_{1_{3,\mathcal{X}}}}}, {\bar{x}^{2t}_{j_{1_{3,\mathcal{X}}}}}) \land \ldots \land 3\text{-REL}_{\mathcal{X}}({\bar{x}}^{3}_{j_{|{\bar{\omega}_{3,{\mathcal{X}}}}|}},{\bar{x}}^{2 t}_{j_{|{\bar{\omega}_{{3,\mathcal{X}}}}|}}) \land \]
\noindent $\langle$\textit{I}$\rangle$, $\langle$\textit{1.I.1}$\rangle$
\[ {\bigwedge_{{\omega}_{2,{\mathcal{Y}_{j_{1_{3,\mathcal{X}}}}}}}} \bigg[ \forall {\bar{y}}^{2t} \bar{y}^{2}_{j_{1_{2,{\mathcal{Y}_{j_{1_{3,\mathcal{X}}}}}}}} \ldots
\bar{y}^{2}_{j_{|{\bar{\omega}}_{2,{\mathcal{Y}_{{j_{1_{3,\mathcal{X}}}}}}}|}} \bigg[ \bigg(\mathrm{X}_{{{{\mathcal{Y}^3}_{j_{1_{3,\mathcal{X}}}}}}, {\omega}_{2,{\mathcal{Y}_{{j_{1_{3,\mathcal{X}}}}}}}} ({\bar{y}}^{2t},\bar{y}^{2}_{j_{1_{2,{\mathcal{Y}_{{j_{1_{3,\mathcal{X}}}}}}}}}, \ldots,\]
\[\bar{y}^{2}_{j_{|{\bar{\omega}}_{2,{\mathcal{Y}_{{j_{1_{3,\mathcal{X}}}}}}}|}})\bigg) \Rightarrow
\exists {\bar{x'}^{2t}} \bar{x}^{2}_{j_{1_{2,{\mathcal{Y}_{{j_{1_{3,\mathcal{X}}}}}}}}} \ldots
\bar{x}^{2}_{j_{|{\bar{\omega}}_{2,{\mathcal{Y}_{{j_{1_{3,\mathcal{X}}}}}}}|}} \bigg(3\text{-REL}_{\mathcal{X}}({\bar{x}^{3}_{j_{1_{3,\mathcal{X}}}}}, {\bar{x'}^{2t}}) \land \]
\[\text{TUPLES-2-REL}_{\mathcal{X},\omega_{2,{\mathcal{Y}_{{j_{1_{3,\mathcal{X}}}}}}}}(\bar{x'}^{2t},\bar{x}^{2}_{j_{1_{2,{\mathcal{Y}_{{j_{1_{3,\mathcal{X}}}}}}}}}, \ldots,
\bar{x}^{2}_{j_{|{\bar{\omega}}_{2,{\mathcal{Y}_{{j_{1_{3,\mathcal{X}}}}}}}|}}) \land \]

\noindent  $\langle$\textit{1(a)i}$\rangle$
\[
\forall \bar{y}^{1t}_{j_{1_{2,{\mathcal{Y}_{{j_{1_{3,\mathcal{X}}}}}}}}} \ldots
\bar{y}^{1t}_{j_{|{\bar{\omega}}_{2,{\mathcal{Y}_{j_{1_{3,\mathcal{X}}}}}}|}} \big[(2\text{-REL}_{{\mathcal{Y}_{{j_{1_{3,\mathcal{X}}}}}}}({\bar{y}^{2}_{j_{1_{2,{\mathcal{Y}_{j_{1_{3,\mathcal{X}}}}}}}}}, {\bar{y}^{1t}_{j_{1_{2,{\mathcal{Y}_{j_{1_{3,\mathcal{X}}}}}}}}}) \Rightarrow \exists {\bar{x}^{1t}_{j_{1_{2,{\mathcal{Y}_{j_{1_{3,\mathcal{X}}}}}}}}} (\]
\[ 2\text{-REL}_{\mathcal{X}}({\bar{x}^{2}_{j_{1_{2,{\mathcal{Y}_{j_{1_{3,\mathcal{X}}}}}}}}}, {\bar{x}^{1t}_{j_{1_{2,{\mathcal{Y}_{j_{1_{3,\mathcal{X}}}}}}}}}) \land \text{``} {\bar{y}^{1t}_{j_{1_{2,{\mathcal{Y}_{j_{1_{3,\mathcal{X}}}}}}}}} = {\bar{x}^{1t}_{j_{1_{2,{\mathcal{Y}_{j_{1_{3,\mathcal{X}}}}}}}}} \text{''})) \land \ldots \land \]
\noindent $\langle$\textit{1(a)ii}$\rangle$
\[(2\text{-REL}_{{\mathcal{Y}_{j_{1_{3,\mathcal{X}}}}}}({\bar{y}^{2}_{j_{|\bar{\omega}_{2,{\mathcal{Y}_{j_{1_{3,\mathcal{X}}}}}}|}}}, {\bar{y}^{1t}_{j_{|\bar{\omega}_{2,{\mathcal{Y}_{j_{1_{3,\mathcal{X}}}}}}|}}}) \Rightarrow \exists {\bar{x}^{1t}_{j_{|\bar{\omega}_{2,{\mathcal{Y}_{j_{1_{3,\mathcal{X}}}}}}|}}} (\]
\[
2\text{-REL}_{\mathcal{X}}({\bar{x}^{2}_{j_{|\bar{\omega}_{2,{\mathcal{Y}_{j_{1_{3,\mathcal{X}}}}}}|}}}, {\bar{x}^{1t}_{j_{|\bar{\omega}_{2,{\mathcal{Y}_{j_{1_{3,\mathcal{X}}}}}}|}}}) \land \text{``} {\bar{y}^{1t}_{j_{|\bar{\omega}_{2,{\mathcal{Y}_{j_{1_{3,\mathcal{X}}}}}}|}}} = {\bar{x}^{1t}_{j_{|\bar{\omega}_{2,{\mathcal{Y}_{j_{1_{3,\mathcal{X}}}}}}|}}} \text{''}))\big ] \land \]
\noindent $\langle$\textit{1(b)i}$\rangle$
\[
 \forall \bar{x'}^{1t}_{j_{1_{2,{\mathcal{Y}_{j_{1_{3,\mathcal{X}}}}}}}} \ldots
\bar{x'}^{1t}_{j_{|{\bar{\omega}}_{2,{{\mathcal{Y}_{j_{1_{3,\mathcal{X}}}}}}}|}} \big[(2\text{-REL}_{\mathcal{X}}({\bar{x}^{2}_{j_{1_{2,{\mathcal{Y}_{j_{1_{3,\mathcal{X}}}}}}}}}, {\bar{x'}^{1t}_{j_{1_{2,{\mathcal{Y}_{j_{1_{3,\mathcal{X}}}}}}}}}) \Rightarrow \exists {\bar{y'}^{1t}_{j_{1_{2,{\mathcal{Y}_{j_{1_{3,\mathcal{X}}}}}}}}} (\]
\[ 2\text{-REL}_{{\mathcal{Y}_{j_{1_{3,\mathcal{X}}}}}}({\bar{y}^{2}_{j_{1_{2,{\mathcal{Y}_{j_{1_{3,\mathcal{X}}}}}}}}}, {\bar{y'}^{1t}_{j_{1_{2,{\mathcal{Y}_{j_{1_{3,\mathcal{X}}}}}}}}}) \land \text{``} {\bar{y'}^{1t}_{j_{1_{2,{\mathcal{Y}_{j_{1_{3,\mathcal{X}}}}}}}}} = {\bar{x'}^{1t}_{j_{1_{2,{\mathcal{Y}_{j_{1_{3,\mathcal{X}}}}}}}}} \text{''})) \land \ldots \land \]
\noindent $\langle$\textit{1(b)ii}$\rangle$
\[(2\text{-REL}_{\mathcal{X}}({\bar{x}^{2}_{j_{|\bar{\omega}_{2,{\mathcal{Y}_{j_{1_{3,\mathcal{X}}}}}}|}}}, {\bar{x'}^{1t}_{j_{|\bar{\omega}_{2,{\mathcal{Y}_{j_{1_{3,\mathcal{X}}}}}}|}}}) \Rightarrow \exists {\bar{y'}^{1t}_{j_{|\bar{\omega}_{2,{\mathcal{Y}_{j_{1_{3,\mathcal{X}}}}}}|}}} (\]
\[
2\text{-REL}_{{\mathcal{Y}_{j_{1_{3,\mathcal{X}}}}}}({\bar{y}^{2}_{j_{|\bar{\omega}_{2,{\mathcal{Y}_{j_{1_{3,\mathcal{X}}}}}}|}}}, {\bar{y'}^{1t}_{j_{|\bar{\omega}_{2,{\mathcal{Y}_{j_{1_{3,\mathcal{X}}}}}}|}}}) \land \text{``} {\bar{y'}^{1t}_{j_{|\bar{\omega}_{2,{\mathcal{Y}_{j_{1_{3,\mathcal{X}}}}}}|}}} = {\bar{x'}^{1t}_{j_{|\bar{\omega}_{2,{\mathcal{Y}_{j_{1_{3,\mathcal{X}}}}}}|}}} \text{''}))\big ] \bigg)\bigg] \land
\]
\noindent $\langle$\textit{2.I.2}$\rangle$
\[\forall \bar{x'}^{2t} \bar{x}^{2}_{j_{1_{2,{\mathcal{Y}_{{j_{1_{3,\mathcal{X}}}}}}}}} \ldots
 \bar{x}^{2}_{j_{|{\bar{\omega}}_{2,{\mathcal{Y}_{{j_{1_{3,\mathcal{X}}}}}}}|}} \bigg[\bigg(3\text{-REL}_{\mathcal{X}}({\bar{x}^{3}_{j_{1_{3,\mathcal{X}}}}}, {\bar{x'}^{2t}}) \land \]
\[
\text{TUPLES-2-REL}_{\mathcal{X},\omega_{2,{\mathcal{Y}_{{j_{1_{3,\mathcal{X}}}}}}}}(\bar{x'}^{2t}, \bar{x}^{2}_{j_{1_{2,{\mathcal{Y}_{{j_{1_{3,\mathcal{X}}}}}}}}}, \ldots,
 \bar{x}^{2}_{j_{|{\bar{\omega}}_{2,{\mathcal{Y}_{{j_{1_{3,\mathcal{X}}}}}}}|}})\bigg) \Rightarrow \]
\[ \exists \bar{y}^{2t} \bar{y}^{2}_{j_{1_{2,{\mathcal{Y}_{{j_{1_{3,\mathcal{X}}}}}}}}} \ldots
 \bar{y}^{2}_{j_{|{\bar{\omega}}_{2,{\mathcal{Y}_{{j_{1_{3,\mathcal{X}}}}}}}|}} \bigg(\mathrm{X}_{{{{\mathcal{Y}^3}_{{j_{1_{3,\mathcal{X}}}}}}},{\omega}_{2,{\mathcal{Y}_{{j_{1_{3,\mathcal{X}}}}}}}}(\bar{y}^{2t},\bar{y}^{2}_{j_{1_{2,{\mathcal{Y}_{{j_{1_{3,\mathcal{X}}}}}}}}}, \ldots, \]
\[ \bar{y}^{2}_{j_{|{\bar{\omega}}_{2,{\mathcal{Y}_{{j_{1_{3,\mathcal{X}}}}}}}|}}) \land \]
\noindent $\langle$\textit{2(a)i}$\rangle$
\[
\forall \bar{y}^{1t}_{j_{1_{2,{\mathcal{Y}_{{j_{1_{3,\mathcal{X}}}}}}}}}  \ldots
\bar{y}^{1t}_{j_{|{\bar{\omega}}_{2,{\mathcal{Y}_{{j_{1_{3,\mathcal{X}}}}}}}|}} \big[(2\text{-REL}_{\mathcal{Y}_{{j_{1_{3,\mathcal{X}}}}}}({\bar{y}^{2}_{j_{1_{2,{\mathcal{Y}_{{j_{1_{3,\mathcal{X}}}}}}}}}}, {\bar{y}^{1t}_{j_{1_{2,{\mathcal{Y}_{{j_{1_{3,\mathcal{X}}}}}}}}}}) \Rightarrow \exists {\bar{x}^{1t}_{j_{1_{2,{\mathcal{Y}_{{j_{1_{3,\mathcal{X}}}}}}}}}} (\]
\[ 2\text{-REL}_{\mathcal{X}}({\bar{x}^{2}_{j_{1_{2,{\mathcal{Y}_{{j_{1_{3,\mathcal{X}}}}}}}}}}, {\bar{x}^{1t}_{j_{1_{2,{\mathcal{Y}_{{j_{1_{3,\mathcal{X}}}}}}}}}}) \land \text{``} {\bar{y}^{1t}_{j_{1_{2,{\mathcal{Y}_{{j_{1_{3,\mathcal{X}}}}}}}}}} = {\bar{x}^{1t}_{j_{1_{2,{\mathcal{Y}_{{j_{1_{3,\mathcal{X}}}}}}}}}} \text{''})) \land \ldots \land \]
\noindent $\langle$\textit{2(a)ii}$\rangle$
\[(2\text{-REL}_{\mathcal{Y}_{{j_{1_{3,\mathcal{X}}}}}}({\bar{y}^{2}_{j_{|\bar{\omega}_{2,{\mathcal{Y}_{{j_{1_{3,\mathcal{X}}}}}}}|}}}, {\bar{y}^{1t}_{j_{|\bar{\omega}_{2,{\mathcal{Y}_{{j_{1_{3,\mathcal{X}}}}}}}|}}}) \Rightarrow \exists {\bar{x}^{1t}_{j_{|\bar{\omega}_{2,{\mathcal{Y}_{{j_{1_{3,\mathcal{X}}}}}}}|}}} (\]
\[2\text{-REL}_{\mathcal{X}}({\bar{x}^{2}_{j_{|\bar{\omega}_{2,{\mathcal{Y}_{{j_{1_{3,\mathcal{X}}}}}}}|}}}, {\bar{x}^{1t}_{j_{|\bar{\omega}_{2,{\mathcal{Y}_{{j_{1_{3,\mathcal{X}}}}}}}|}}}) \land \text{``} {\bar{y}^{1t}_{j_{|\bar{\omega}_{2,{\mathcal{Y}_{{j_{1_{3,\mathcal{X}}}}}}}|}}} = {\bar{x}^{1t}_{j_{|\bar{\omega}_{2,{\mathcal{Y}_{{j_{1_{3,\mathcal{X}}}}}}}|}}} \text{''}))\big] \land \]
\noindent $\langle$\textit{2(b)i}$\rangle$
\[
 \forall \bar{x'}^{1t}_{j_{1_{2,{\mathcal{Y}_{{j_{1_{3,\mathcal{X}}}}}}}}} \ldots
\bar{x'}^{1t}_{j_{|{\bar{\omega}}_{2,{\mathcal{Y}_{{j_{1_{3,\mathcal{X}}}}}}}|}} \big[(2\text{-REL}_{\mathcal{X}}({\bar{x}^{2}_{j_{1_{2,{\mathcal{Y}_{{j_{1_{3,\mathcal{X}}}}}}}}}}, {\bar{x'}^{1t}_{j_{1_{2,{\mathcal{Y}_{{j_{1_{3,\mathcal{X}}}}}}}}}}) \Rightarrow \exists {\bar{y'}^{1t}_{j_{1_{2,{\mathcal{Y}_{{j_{1_{3,\mathcal{X}}}}}}}}}} (\]
\[ 2\text{-REL}_{\mathcal{Y}_{{j_{1_{3,\mathcal{X}}}}}}({\bar{y}^{2}_{j_{1_{2,{\mathcal{Y}_{{j_{1_{3,\mathcal{X}}}}}}}}}}, {\bar{y'}^{1t}_{j_{1_{2,{\mathcal{Y}_{{j_{1_{3,\mathcal{X}}}}}}}}}}) \land \text{``} {\bar{y'}^{1t}_{j_{1_{2,{\mathcal{Y}_{{j_{1_{3,\mathcal{X}}}}}}}}}} = {\bar{x'}^{1t}_{j_{1_{2,{\mathcal{Y}_{{j_{1_{3,\mathcal{X}}}}}}}}}} \text{''})) \land \ldots \land \]
\noindent $\langle$\textit{2(b)ii}$\rangle$
\[(2\text{-REL}_{\mathcal{X}}({\bar{x}^{2}_{j_{|\bar{\omega}_{2,{\mathcal{Y}_{{j_{1_{3,\mathcal{X}}}}}}}|}}}, {\bar{x'}^{1t}_{j_{|\bar{\omega}_{2,{\mathcal{Y}_{{j_{1_{3,\mathcal{X}}}}}}}|}}}) \Rightarrow \exists {\bar{y'}^{1t}_{j_{|\bar{\omega}_{2,{\mathcal{Y}_{{j_{1_{3,\mathcal{X}}}}}}}|}}} (\]
\[
2\text{-REL}_{\mathcal{Y}_{{j_{1_{3,\mathcal{X}}}}}}({\bar{y}^{2}_{j_{|\bar{\omega}_{2,{\mathcal{Y}_{{j_{1_{3,\mathcal{X}}}}}}}|}}}, {\bar{y'}^{1t}_{j_{|\bar{\omega}_{2,{\mathcal{Y}_{{j_{1_{3,\mathcal{X}}}}}}}|}}}) \land \text{``} {\bar{y'}^{1t}_{j_{|\bar{\omega}_{2,{\mathcal{Y}_{{j_{1_{3,\mathcal{X}}}}}}}|}}} = {\bar{x'}^{1t}_{j_{|\bar{\omega}_{2,{\mathcal{Y}_{{j_{1_{3,\mathcal{X}}}}}}}|}}} \text{''}))\big ] \bigg) \bigg]\bigg] \land \]
\noindent $\langle$\textit{II}$\rangle$,  $\langle$\textit{1.II.1}$\rangle$
\[\ldots \land {\bigwedge_{{\omega}_{2,{\mathcal{Y}_{{j_{|{\bar{\omega}}_{3,\mathcal{X}}|}}}}}}} \bigg[  \forall {\bar{y}}^{2t} \bar{y}^{2}_{j_{1_{2,{\mathcal{Y}_{{j_{|\bar{\omega}_{3,\mathcal{X}}|}}}}}}} \ldots  \bar{y}^{2}_{j_{|{\bar{\omega}}_{2,{\mathcal{Y}_{{j_{|\bar{\omega}_{3,\mathcal{X}}|}}}}}|}} \bigg[ \bigg(\mathrm{X}_{{{{\mathcal{Y}^3}_{{j_{|\bar{\omega}_{3,\mathcal{X}}|}}}}} ,{\omega}_{2,{\mathcal{Y}_{{j_{|\bar{\omega}_{3,\mathcal{X}}|}}}}}} ({\bar{y}}^{2t},\bar{y}^{2}_{j_{1_{2,{\mathcal{Y}_{{j_{|\bar{\omega}_{3,\mathcal{X}}|}}}}}}}, \]
\[\ldots,  \bar{y}^{2}_{j_{|{\bar{\omega}}_{2,{\mathcal{Y}_{{j_{|\bar{\omega}_{3,\mathcal{X}}|}}}}}|}}) \bigg)\Rightarrow \exists {\bar{x'}^{2t}}  \bar{x}^{2}_{j_{1_{2,{\mathcal{Y}_{{j_{|\bar{\omega}_{3,\mathcal{X}}|}}}}}}} \ldots  \bar{x}^{2}_{j_{|{\bar{\omega}}_{2,{\mathcal{Y}_{{j_{|\bar{\omega}_{3,\mathcal{X}}|}}}}}|}} \bigg(3\text{-REL}_{\mathcal{X}}(\bar{x}^{3}_{j_{|\bar{\omega}_{3,\mathcal{X}}|}}, {\bar{x'}^{2t}}) \land \]
\[\text{TUPLES-2-REL}_{\mathcal{X},\omega_{2,{\mathcal{Y}_{{j_{|{\bar{\omega}}_{3,\mathcal{X}}|}}}}}}(\bar{x'}^{2t}, \bar{x}^{2}_{j_{1_{2,{\mathcal{Y}_{{j_{|\bar{\omega}_{3,\mathcal{X}}|}}}}}}}, \ldots,  \bar{x}^{2}_{j_{|{\bar{\omega}}_{2,{\mathcal{Y}_{{j_{|\bar{\omega}_{3,\mathcal{X}}|}}}}}|}} ) \land \]
\noindent $\langle$\textit{1(a)i}$\rangle$
\[
\forall \bar{y}^{1t}_{j_{1_{2,{\mathcal{Y}_{{j_{|\bar{\omega}_{3,\mathcal{X}}|}}}}}}} \ldots
\bar{y}^{1t}_{j_{|{\bar{\omega}}_{2,{\mathcal{Y}_{{j_{|\bar{\omega}_{3,\mathcal{X}}|}}}}}|}} \big[(2\text{-REL}_{{\mathcal{Y}_{{j_{|\bar{\omega}_{3,\mathcal{X}}|}}}}}({\bar{y}^{2}_{j_{1_{2,{\mathcal{Y}_{{j_{|\bar{\omega}_{3,\mathcal{X}}|}}}}}}}}, {\bar{y}^{1t}_{j_{1_{2,{\mathcal{Y}_{{j_{|\bar{\omega}_{3,\mathcal{X}}|}}}}}}}}) \Rightarrow \exists {\bar{x}^{1t}_{j_{1_{2,{\mathcal{Y}_{{j_{|\bar{\omega}_{3,\mathcal{X}}|}}}}}}}} (\]
\[ 2\text{-REL}_{\mathcal{X}}({\bar{x}^{2}_{j_{1_{2,{\mathcal{Y}_{{j_{|\bar{\omega}_{3,\mathcal{X}}|}}}}}}}}, {\bar{x}^{1t}_{j_{1_{2,{\mathcal{Y}_{{j_{|\bar{\omega}_{3,\mathcal{X}}|}}}}}}}}) \land \text{``} {\bar{y}^{1t}_{j_{1_{2,{\mathcal{Y}_{{j_{|\bar{\omega}_{3,\mathcal{X}}|}}}}}}}} = {\bar{x}^{1t}_{j_{1_{2,{\mathcal{Y}_{{j_{|\bar{\omega}_{3,\mathcal{X}}|}}}}}}}} \text{''})) \land \ldots \land \]
\noindent $\langle$\textit{1(a)ii}$\rangle$
\[(2\text{-REL}_{\mathcal{Y}_{{j_{|\bar{\omega}_{3,\mathcal{X}}|}}}}({\bar{y}^{2}_{j_{|\bar{\omega}_{2,{\mathcal{Y}_{{j_{|\bar{\omega}_{3,\mathcal{X}}|}}}}}|}}}, {\bar{y}^{1t}_{j_{|\bar{\omega}_{2,{\mathcal{Y}_{{j_{|\bar{\omega}_{3,\mathcal{X}}|}}}}}|}}}) \Rightarrow \exists {\bar{x}^{1t}_{j_{|\bar{\omega}_{2,{\mathcal{Y}_{{j_{|\bar{\omega}_{3,\mathcal{X}}|}}}}}|}}} (\]
\[
2\text{-REL}_{\mathcal{X}}({\bar{x}^{2}_{j_{|\bar{\omega}_{2,{\mathcal{Y}_{{j_{|\bar{\omega}_{3,\mathcal{X}}|}}}}}|}}}, {\bar{x}^{1t}_{j_{|\bar{\omega}_{2,{\mathcal{Y}_{{j_{|\bar{\omega}_{3,\mathcal{X}}|}}}}}|}}}) \land \text{``} {\bar{y}^{1t}_{j_{|\bar{\omega}_{2,{\mathcal{Y}_{{j_{|\bar{\omega}_{3,\mathcal{X}}|}}}}}|}}} = {\bar{x}^{1t}_{j_{|\bar{\omega}_{2,{\mathcal{Y}_{{j_{|\bar{\omega}_{3,\mathcal{X}}|}}}}}|}}} \text{''}))\big ] \land \]
\noindent $\langle$\textit{1(b)i}$\rangle$
\[
 \forall \bar{x'}^{1t}_{j_{1_{2,{\mathcal{Y}_{{j_{|\bar{\omega}_{3,\mathcal{X}}|}}}}}}} \ldots
\bar{x'}^{1t}_{j_{|{\bar{\omega}}_{2,{\mathcal{Y}_{{j_{|\bar{\omega}_{3,\mathcal{X}}|}}}}}|}} \big[(2\text{-REL}_{\mathcal{X}}({\bar{x}^{2}_{j_{1_{2,{\mathcal{Y}_{{j_{|\bar{\omega}_{3,\mathcal{X}}|}}}}}}}}, {\bar{x'}^{1t}_{j_{1_{2,{\mathcal{Y}_{{j_{|\bar{\omega}_{3,\mathcal{X}}|}}}}}}}}) \Rightarrow \exists {\bar{y'}^{1t}_{j_{1_{2,{\mathcal{Y}_{{j_{|\bar{\omega}_{3,\mathcal{X}}|}}}}}}}} (\]
\[ 2\text{-REL}_{{\mathcal{Y}_{{j_{|\bar{\omega}_{3,\mathcal{X}}|}}}}}({\bar{y}^{2}_{j_{1_{2,{\mathcal{Y}_{{j_{|\bar{\omega}_{3,\mathcal{X}}|}}}}}}}}, {\bar{y'}^{1t}_{j_{1_{2,{\mathcal{Y}_{{j_{|\bar{\omega}_{3,\mathcal{X}}|}}}}}}}}) \land \text{``} {\bar{y'}^{1t}_{j_{1_{2,{\mathcal{Y}_{{j_{|\bar{\omega}_{3,\mathcal{X}}|}}}}}}}} = {\bar{x'}^{1t}_{j_{1_{2,{\mathcal{Y}_{{j_{|\bar{\omega}_{3,\mathcal{X}}|}}}}}}}} \text{''})) \land \ldots \land \]
\noindent $\langle$\textit{1(b)ii}$\rangle$
\[(2\text{-REL}_{\mathcal{X}}({\bar{x}^{2}_{j_{|\bar{\omega}_{2,{\mathcal{Y}_{{j_{|\bar{\omega}_{3,\mathcal{X}}|}}}}}|}}}, {\bar{x'}^{1t}_{j_{|\bar{\omega}_{2,{\mathcal{Y}_{{j_{|\bar{\omega}_{3,\mathcal{X}}|}}}}}|}}}) \Rightarrow \exists {\bar{y'}^{1t}_{j_{|\bar{\omega}_{2,{\mathcal{Y}_{{j_{|\bar{\omega}_{3,\mathcal{X}}|}}}}}|}}} (\]
\[
2\text{-REL}_{{\mathcal{Y}_{{j_{|\bar{\omega}_{3,\mathcal{X}}|}}}}}({\bar{y}^{2}_{j_{|\bar{\omega}_{2,{\mathcal{Y}_{{j_{|\bar{\omega}_{3,\mathcal{X}}|}}}}}|}}}, {\bar{y'}^{1t}_{j_{|\bar{\omega}_{2,{\mathcal{Y}_{{j_{|\bar{\omega}_{3,\mathcal{X}}|}}}}}|}}}) \land \text{``} {\bar{y'}^{1t}_{j_{|\bar{\omega}_{2,{\mathcal{Y}_{{j_{|\bar{\omega}_{3,\mathcal{X}}|}}}}}|}}} = {\bar{x'}^{1t}_{j_{|\bar{\omega}_{2,{\mathcal{Y}_{{j_{|\bar{\omega}_{3,\mathcal{X}}|}}}}}|}}} \text{''}))\big ] \bigg) \bigg]  \land
 \]
 \noindent $\langle$\textit{2.II.2}$\rangle$
\[\forall \bar{x'}^{2t} \bar{x}^{2}_{j_{1_{2,{\mathcal{Y}_{{j_{|{\bar{\omega}}_{3,\mathcal{X}}|}}}}}}} \ldots
 \bar{x}^{2}_{j_{|{\bar{\omega}}_{2,{\mathcal{Y}_{j_{|{\bar{\omega}}_{3,\mathcal{X}}|}}}}|}} \bigg[\bigg(3\text{-REL}_{\mathcal{X}}({\bar{x}^{3}_{j_{|{\bar{\omega}}_{3,\mathcal{X}}|}}}, {\bar{x'}^{2t}}) \land \]
\[
\text{TUPLES-2-REL}_{\mathcal{X},\omega_{2,{\mathcal{Y}_{{j_{|{\bar{\omega}}_{3,\mathcal{X}}|}}}}}}(\bar{x'}^{2t}, \bar{x}^{2}_{j_{1_{2,{\mathcal{Y}_{j_{|{\bar{\omega}}_{3,\mathcal{X}}|}}}}}}, \ldots,
 \bar{x}^{2}_{j_{|{\bar{\omega}}_{2,{\mathcal{Y}_{{j_{|{\bar{\omega}}_{3,\mathcal{X}}|}}}}}|}})\bigg) \Rightarrow \]
\[ \exists \bar{y}^{2t} \bar{y}^{2}_{j_{1_{2,{\mathcal{Y}_{{j_{|{\bar{\omega}}_{3,\mathcal{X}}|}}}}}}} \ldots
 \bar{y}^{2}_{j_{|{\bar{\omega}}_{2,{\mathcal{Y}_{{j_{|{\bar{\omega}}_{3,\mathcal{X}}|}}}}}|}} \bigg(\mathrm{X}_{{{{\mathcal{Y}^3}_{{j_{|{\bar{\omega}}_{3,\mathcal{X}}|}}}}},{\omega}_{2,{\mathcal{Y}_{{j_{|{\bar{\omega}}_{3,\mathcal{X}}|}}}}}}(\bar{y}^{2t},\bar{y}^{2}_{j_{1_{2,{\mathcal{Y}_{{j_{|{\bar{\omega}}_{3,\mathcal{X}}|}}}}}}}, \ldots, \]
\[ \bar{y}^{2}_{j_{|{\bar{\omega}}_{2,{\mathcal{Y}_{{j_{|{\bar{\omega}}_{3,\mathcal{X}}|}}}}}|}}) \land \]
\noindent $\langle$\textit{2(a)i}$\rangle$
\[
\forall \bar{y}^{1t}_{j_{1_{2,{\mathcal{Y}_{{j_{|{\bar{\omega}}_{3,\mathcal{X}}|}}}}}}}  \ldots
\bar{y}^{1t}_{j_{|{\bar{\omega}}_{2,{\mathcal{Y}_{{j_{|{\bar{\omega}}_{3,\mathcal{X}}|}}}}}|}} \big[(2\text{-REL}_{\mathcal{Y}_{{j_{|{\bar{\omega}}_{3,\mathcal{X}}|}}}}({\bar{y}^{2}_{j_{1_{2,{\mathcal{Y}_{{j_{|{\bar{\omega}}_{3,\mathcal{X}}|}}}}}}}}, {\bar{y}^{1t}_{j_{1_{2,{\mathcal{Y}_{{j_{|{\bar{\omega}}_{3,\mathcal{X}}|}}}}}}}}) \Rightarrow \exists {\bar{x}^{1t}_{j_{1_{2,{\mathcal{Y}_{{j_{|{\bar{\omega}}_{3,\mathcal{X}}|}}}}}}}} (\]
\[ 2\text{-REL}_{\mathcal{X}}({\bar{x}^{2}_{j_{1_{2,{\mathcal{Y}_{{j_{|{\bar{\omega}}_{3,\mathcal{X}}|}}}}}}}}, {\bar{x}^{1t}_{j_{1_{2,{\mathcal{Y}_{{j_{|{\bar{\omega}}_{3,\mathcal{X}}|}}}}}}}}) \land \text{``} {\bar{y}^{1t}_{j_{1_{2,{\mathcal{Y}_{j_{|{\bar{\omega}}_{3,\mathcal{X}}|}}}}}}} = {\bar{x}^{1t}_{j_{1_{2,{\mathcal{Y}_{j_{|{\bar{\omega}}_{3,\mathcal{X}}|}}}}}}} \text{''})) \land \ldots \land \]
\noindent $\langle$\textit{2(a)ii}$\rangle$
\[(2\text{-REL}_{\mathcal{Y}_{{j_{|{\bar{\omega}}_{3,\mathcal{X}}|}}}}({\bar{y}^{2}_{j_{|\bar{\omega}_{2,{\mathcal{Y}_{{j_{|{\bar{\omega}}_{3,\mathcal{X}}|}}}}}|}}}, {\bar{y}^{1t}_{j_{|\bar{\omega}_{2,{\mathcal{Y}_{{j_{|{\bar{\omega}}_{3,\mathcal{X}}|}}}}}|}}}) \Rightarrow \exists {\bar{x}^{1t}_{j_{|\bar{\omega}_{2,{\mathcal{Y}_{{j_{|{\bar{\omega}}_{3,\mathcal{X}}|}}}}}|}}} (\]
\[2\text{-REL}_{\mathcal{X}}({\bar{x}^{2}_{j_{|\bar{\omega}_{2,{\mathcal{Y}_{{j_{|{\bar{\omega}}_{3,\mathcal{X}}|}}}}}|}}}, {\bar{x}^{1t}_{j_{|\bar{\omega}_{2,{\mathcal{Y}_{{j_{|{\bar{\omega}}_{3,\mathcal{X}}|}}}}}|}}}) \land \text{``} {\bar{y}^{1t}_{j_{|\bar{\omega}_{2,{\mathcal{Y}_{{j_{|{\bar{\omega}}_{3,\mathcal{X}}|}}}}}|}}} = {\bar{x}^{1t}_{j_{|\bar{\omega}_{2,{\mathcal{Y}_{{j_{|{\bar{\omega}}_{3,\mathcal{X}}|}}}}}|}}} \text{''}))\big] \land \]
\noindent $\langle$\textit{2(b)i}$\rangle$
\[
 \forall \bar{x'}^{1t}_{j_{1_{2,{\mathcal{Y}_{{j_{|{\bar{\omega}}_{3,\mathcal{X}}|}}}}}}} \ldots
\bar{x'}^{1t}_{j_{|{\bar{\omega}}_{2,{\mathcal{Y}_{{j_{|{\bar{\omega}}_{3,\mathcal{X}}|}}}}}|}} \big[(2\text{-REL}_{\mathcal{X}}({\bar{x}^{2}_{j_{1_{2,{\mathcal{Y}_{{j_{|{\bar{\omega}}_{3,\mathcal{X}}|}}}}}}}}, {\bar{x'}^{1t}_{j_{1_{2,{\mathcal{Y}_{{j_{|{\bar{\omega}}_{3,\mathcal{X}}|}}}}}}}}) \Rightarrow \exists {\bar{y'}^{1t}_{j_{1_{2,{\mathcal{Y}_{{j_{|{\bar{\omega}}_{3,\mathcal{X}}|}}}}}}}} (\]
\[ 2\text{-REL}_{\mathcal{Y}_{{j_{|{\bar{\omega}}_{3,\mathcal{X}}|}}}}({\bar{y}^{2}_{j_{1_{2,{\mathcal{Y}_{{j_{|{\bar{\omega}}_{3,\mathcal{X}}|}}}}}}}}, {\bar{y'}^{1t}_{j_{1_{2,{\mathcal{Y}_{{j_{|{\bar{\omega}}_{3,\mathcal{X}}|}}}}}}}}) \land \text{``} {\bar{y'}^{1t}_{j_{1_{2,{\mathcal{Y}_{{j_{|{\bar{\omega}}_{3,\mathcal{X}}|}}}}}}}} = {\bar{x'}^{1t}_{j_{1_{2,{\mathcal{Y}_{{j_{|{\bar{\omega}}_{3,\mathcal{X}}|}}}}}}}} \text{''})) \land \ldots \land \]
\noindent $\langle$\textit{2(b)ii}$\rangle$
\[(2\text{-REL}_{\mathcal{X}}({\bar{x}^{2}_{j_{|\bar{\omega}_{2,{\mathcal{Y}_{{j_{|{\bar{\omega}}_{3,\mathcal{X}}|}}}}}|}}}, {\bar{x'}^{1t}_{j_{|\bar{\omega}_{2,{\mathcal{Y}_{{j_{|{\bar{\omega}}_{3,\mathcal{X}}|}}}}}|}}}) \Rightarrow \exists {\bar{y'}^{1t}_{j_{|\bar{\omega}_{2,{\mathcal{Y}_{{j_{|{\bar{\omega}}_{3,\mathcal{X}}|}}}}}|}}} (\]
\[
2\text{-REL}_{\mathcal{Y}_{{j_{|{\bar{\omega}}_{3,\mathcal{X}}|}}}}({\bar{y}^{2}_{j_{|\bar{\omega}_{2,{\mathcal{Y}_{{j_{|{\bar{\omega}}_{3,\mathcal{X}}|}}}}}|}}}, {\bar{y'}^{1t}_{j_{|\bar{\omega}_{2,{\mathcal{Y}_{{j_{|{\bar{\omega}}_{3,\mathcal{X}}|}}}}}|}}}) \land \text{``} {\bar{y'}^{1t}_{j_{|\bar{\omega}_{2,{\mathcal{Y}_{{j_{|{\bar{\omega}}_{3,\mathcal{X}}|}}}}}|}}} = {\bar{x'}^{1t}_{j_{|\bar{\omega}_{2,{\mathcal{Y}_{{j_{|{\bar{\omega}}_{3,\mathcal{X}}|}}}}}|}}} \text{''}))\big ] \bigg) \bigg] \bigg]\bigg) \]

\section{The case of Existential Formulae in $\mathrm{HO}^{4,P}$}\label{sectionB}

The existential case $\exists \mathcal{X}^{4,d,\tau}(\mathcal{\varphi})$ with $|\tau|=s$, downward polynomially bounded, with degree $d\ge 1$ is as follows.

\[
\exists \{X^{d+|\bar{f}_{\bar{\omega}_{3,\mathcal{X}}}|}_{\mathcal{X}^4,\omega_{3,\mathcal{X}}}\}_{\omega_{3,\mathcal{X}}} \;\exists 3\text{-REL}_{\mathcal{X}}^{2d} \;\exists \{\text{TUPLES-2-REL}_{\mathcal{X},\omega_{2,\mathcal{X}}}^{d+|\bar{f}_{\bar{\omega}_{2,\mathcal{X}}}|}\}_{\omega_{2,\mathcal{X}}}\]
\[
\exists 2\text{-REL}_{\mathcal{X}}^{d+s}\bigg[(\text{``database $\mathcal{X}^4$ has referential integrity''}) \land \]
\[
\bigg[\forall \bar{x}^{3t} \bigg[\bigwedge_{\omega_{3,\mathcal{X}}} \forall  \bar{f}_{{\bar{\omega}}_{3,\mathcal{X}}} \bigg[\mathrm{X}_{\mathcal{X}^4,\omega_{3,\mathcal{X}}}(\bar{x}^{3t},\bar{f}_{\bar{\omega}_{3,\mathcal{X}}})  \Rightarrow\]
\[
 \bigg(\bigwedge_{{\omega'}_{3,\mathcal{X}}\neq \omega_{3,\mathcal{X}}} \forall  \bar{f'}_{\bar{\omega}^\prime_{3,\mathcal{X}}} \bigg( \neg \mathrm{X}_{\mathcal{X}^4,\omega'_{3,\mathcal{X}}}(\bar{x}^{3t},\bar{f'}_{\bar{\omega'}_{3,\mathcal{X}}})\bigg)\bigg)\bigg]\bigg]\bigg]
\land \]
\[
\bigg[\forall \bar{x}^{2t} \bigg[\bigwedge_{\omega_{2,\mathcal{X}}} \forall \bar{f}_{{\bar{\omega}}_{2,\mathcal{X}}}\bigg[\text{TUPLES-2-REL}_{\mathcal{X},\omega_{2,\mathcal{X}}}(\bar{x}^{2t},\bar{f}_{\bar{\omega}_{2,\mathcal{X}}}) \Rightarrow \]
\[
 \bigg(\bigwedge_{{\omega'}_{2,\mathcal{X}}\neq \omega_{2,\mathcal{X}}} \forall \bar{f}^\prime_{{\bar{\omega}}^\prime_{2,\mathcal{X}}} \bigg(\neg \text{TUPLES-2-REL}_{\mathcal{X},\omega'_{2,\mathcal{X}}}(\bar{x}^{2t},\bar{f'}_{\bar{\omega'}_{2,\mathcal{X}}})\bigg)\bigg)\bigg]\bigg]\bigg] \land \hat{\varphi}\bigg],
 \]
where $\hat{\varphi}$ is the $\text{HO}^{4,P}$ formula $\varphi$, obtained by inductively applying the translations described above.

\vspace*{1cm}

\noindent``database $\mathcal{X}^4$ has referential integrity'':

\noindent``in $\mathrm{X}_{\mathcal{X}^4,\omega_{3,\mathcal{X}}}$ there are no two tuples of TO relations with the same id of TO relations tuple'':

\[
\bigg[ \bigwedge_{\omega_{3,\mathcal{X}}} \forall \bar{x}^{3t} \neg \exists \bar{x}^{3}_{j_{1_{3,\mathcal{X}}}} \bar{x'}^{3}_{j_{1_{3,\mathcal{X}}}} \ldots \bar{x}^{3}_{j_{|{\bar{\omega}}_{3,\mathcal{X}}|}} \bar{x'}^{3}_{j_{|{\bar{\omega}}_{3,\mathcal{X}}|}} \bigg((\bar{x}^3_{j_{1_{3,\mathcal{X}}}} \neq \bar{x'}^{3}_{j_{1_{3,\mathcal{X}}}} \lor \ldots \]
\[ \lor \bar{x}^{3}_{j_{|{\bar{\omega}}_{3,\mathcal{X}}|}} \neq \bar{x'}^{3}_{j_{|{\bar{\omega}}_{3,\mathcal{X}}|}}) \land \mathrm{X}_{\mathcal{X}^4,\omega_{3,\mathcal{X}}}(\bar{x}^{3t},\bar{x}^3_{j_{1_{3,\mathcal{X}}}},\ldots,\bar{x}^3_{j_{|\bar{\omega}_{3,\mathcal{X}}|}}) \land \]
\[
\mathrm{X}_{\mathcal{X}^4,\omega_{3,\mathcal{X}}}(\bar{x}^{3t},\bar{x'}^3_{j_{1_{3,\mathcal{X}}}},\ldots,\bar{x'}^3_{j_{|\bar{\omega}_{3,\mathcal{X}}|}})\bigg) \land
\]

\noindent``all TO relations in the tuples in $\mathrm{X}_{\mathcal{X}^4,\omega_{3,\mathcal{X}}}$ are in 3-REL$_{\mathcal{X}}$'':

\[
\forall \bar{x}^{3t} \bar{x}^{3}_{j_{1_{3,\mathcal{X}}}} \ldots \bar{x}^{3}_{j_{|{\bar{\omega}}_{3,\mathcal{X}}|}} \bigg(\mathrm{X}_{\mathcal{X}^4,\omega_{3,\mathcal{X}}}(\bar{x}^{3t},\bar{x}^3_{j_{1_{3,\mathcal{X}}}},\ldots,\bar{x}^3_{j_{|{\bar{\omega}}_{3,\mathcal{X}}|}})\Rightarrow \]
\[
\exists \bar{x}^{2t}_{j_{1_{3,\mathcal{X}}}} \ldots \bar{x}^{2t}_{j_{|{\bar{\omega}}_{3,\mathcal{X}}|}}\bigg(3\text{-REL}_{\mathcal{X}}(\bar{x}^3_{j_{1_{3,\mathcal{X}}}},\bar{x}^{2t}_{j_{1_{3,\mathcal{X}}}})\land \ldots \land 3\text{-REL}_{\mathcal{X}}(\bar{x}^3_{j_{|{\bar{\omega}}_{3,\mathcal{X}}|}},\bar{x}^{2t}_{j_{|{\bar{\omega}}_{3,\mathcal{X}}|}})\bigg)\bigg)\bigg] \land
\]

\noindent``every TO relation in 3-REL$_{\mathcal{X}}$ is in some tuple in $\mathrm{X}_{\mathcal{X}^4,\omega_{3,\mathcal{X}}}$'':
\[
\bigg[\forall \bar{x}^3 \bar{x}^{2t} \bigg(3\text{-REL}_{\mathcal{X}}(\bar{x}^3,\bar{x}^{2t}) \Rightarrow \exists \bar{x}^{3t}\bigg( \bigvee_{\omega_{3,\mathcal{X}}} \exists \bar{x}^3_{j_{1_{3,\mathcal{X}}}} \ldots \bar{x}^3_{j_{|\bar{\omega}_{3,\mathcal{X}}|}} ((``\bar{x}^3 = \bar{x}^3_{j_{1_{3,\mathcal{X}}}}\text{''} \lor \ldots \lor \]
\[ ``\bar{x}^3=\bar{x}^3_{j_{|{\bar{\omega}_{3,\mathcal{X}}}|}}\text{''}) \land \mathrm{X}_{\mathcal{X}^4,\omega_{3,\mathcal{X}}} (\bar{x}^{3t},\bar{x}^3_{j_{1_{3,\mathcal{X}}}},\ldots,\bar{x}^3_{j_{|{\bar{\omega}_{3,\mathcal{X}}}|}})) \bigg)\bigg)\bigg]\land\]

\noindent``all tuples of SO relations in 3-REL$_\mathcal{X}$ are in some TUPLES-2-REL$_{\mathcal{X},\omega_{2,\mathcal{X}}}$ '':
\[
\forall \bar{x}^3 \bar{x}^{2t} \bigg[3\text{-REL}_{\mathcal{X}}(\bar{x}^3,\bar{x}^{2t}) \Rightarrow \bigvee_{\omega_{2,\mathcal{X}}} \exists \bar{x}^2_{j_{1_{2,\mathcal{X}}}} \ldots \bar{x}^2_{j_{|{\bar{\omega}}_{2,\mathcal{X}}|}} \bigg(\text{TUPLES-2-REL}_{\mathcal{X},\omega_{2,\mathcal{X}}}(\bar{x}^{2t},\]
\[
\bar{x}^2_{j_{1_{2,\mathcal{X}}}},\ldots,\bar{x}^2_{j_{|{\bar{\omega}}_{2,\mathcal{X}}}|})\bigg)\bigg] \land \]
\noindent``all SO relations in the tuples in TUPLES-2-REL$_{\mathcal{X},\omega_{2,\mathcal{X}}}$  are in 2-REL$_\mathcal{X}$'':
\[
\bigg[ \bigwedge_{\omega_{2,\mathcal{X}}} \forall \bar{x}^{2t}\bar{x}^2_{j_{1_{2,\mathcal{X}}}} \ldots \bar{x}^2_{j_{{|\bar{\omega}}_{2,\mathcal{X}}|}}\bigg[\text{TUPLES-2-REL}_{\mathcal{X},\omega_{2,\mathcal{X}}} (\bar{x}^{2t},\bar{x}^2_{j_{1_{2,\mathcal{X}}}},\ldots, \bar{x}^2_{j_{{|\bar{\omega}}_{2,\mathcal{X}}|}})\Rightarrow \]
\[
\exists \bar{x}^{1t}_{j_{1_{2,\mathcal{X}}}} \ldots \bar{x}^{1t}_{j_{{|\bar{\omega}}_{2,\mathcal{X}}|}}\bigg(2\text{-REL}_{\mathcal{X}}(\bar{x}^2_{j_{1_{2,\mathcal{X}}}},\bar{x}^{1t}_{j_{1_{2,\mathcal{X}}}}) \land \ldots \land 2\text{-REL}_{\mathcal{X}}(\bar{x}^2_{j_{|{\omega}_{2,\mathcal{X}}|}},\bar{x}^{1t}_{j_{|{\omega}_{2,\mathcal{X}}|}})\bigg)\bigg]
\bigg] \land \]

\noindent``every SO relation in 2-REL$_\mathcal{X}$ is in some tuple in TUPLES-2-REL$_{\mathcal{X},\omega_{2,\mathcal{X}}}$ '':
\[
\bigg[\forall \bar{x}^2 \forall\bar{x}^{1t}\bigg[2\text{-REL}_{\mathcal{X}}(\bar{x}^2,\bar{x}^{1t}) \Rightarrow \exists \bar{x}^{2t} \bigg(\bigvee_{\omega_{2,\mathcal{X}}} \exists \bar{x}^2_{j_{1_{2,\mathcal{X}}}}\ldots \bar{x}^2_{j_{|\bar{\omega}_{2,\mathcal{X}}|}} ((``\bar{x}^2=\bar{x}^{2}_{j_{1_{2,\mathcal{X}}}}\text{''} \lor \ldots \lor \]
\[``\bar{x}^2=\bar{x}^2_{j_{|\bar{\omega}_{2,\mathcal{X}}|}}\text{''}) \land  \text{TUPLES-2-REL}_{\mathcal{X},\omega_{2,\mathcal{X}}}(\bar{x}^{2t},\bar{x}^2_{j_{1_{2,\mathcal{X}}}},\ldots,\bar{x}^2_{j_{|\bar{\omega}_{2,\mathcal{X}}|}}))\bigg)\bigg]\bigg] \land
\]

\noindent``in TUPLES-2-REL$_{\mathcal{X},\omega_{2,\mathcal{X}}}$ there are no two tuples of SO relations with the same id of SO relation tuple'':
\[\bigg[\bigwedge_{\omega_{2,\mathcal{X}}} \forall \bar{x}^{2t} \neg \exists \bar{x}^2_{j_{1_{2,\mathcal{X}}}} \bar{x'}^2_{j_{1_{2,\mathcal{X}}}} \ldots \bar{x}^2_{j_{|{\bar\omega}_{2,\mathcal{X}}|}} \bar{x'}^2_{j_{|{\bar\omega}_{2,\mathcal{X}}|}} \bigg((``\bar{x}^{2}_{j_{1_{2,\mathcal{X}}}}\neq\bar{x'}^{2}_{j_{1_{2,\mathcal{X}}}}\text{''} \lor \ldots \lor \]
\[``\bar{x}^2_{j_{|\bar{\omega}_{2,\mathcal{X}}|}}\neq\bar{x'}^2_{j_{|\bar{\omega}_{2,\mathcal{X}}|}}\text{''}) \land  \text{TUPLES-2-REL}_{\mathcal{X},\omega_{2,\mathcal{X}}}(\bar{x}^{2t},\bar{x}^2_{j_{1_{2,\mathcal{X}}}},\ldots,\bar{x}^2_{j_{|\bar{\omega}_{2,\mathcal{X}}|}}) \land \]
\[
\text{TUPLES-2-REL}_{\mathcal{X},\omega_{2,\mathcal{X}}}(\bar{x}^{2t},\bar{x'}^2_{j_{1_{2,\mathcal{X}}}},\ldots,\bar{x'}^2_{j_{|\bar{\omega}_{2,\mathcal{X}}|}})\bigg) \bigg] \land
\]

\noindent``every tuple of SO relations in TUPLES-2-REL$_{\mathcal{X},\omega_{2,\mathcal{X}}}$ is in some TO relation in 3-REL$_{\mathcal{X}}$'':
\[ \bigg[\forall \bar{x}^{2t} \bigwedge_{\omega_{2,\mathcal{X}}} \forall \bar{x}^2_{j_{1_{2,\mathcal{X}}}}\ldots \bar{x}^2_{j_{|{\bar{\omega}_{2,\mathcal{X}}}|}} \bigg[\text{TUPLES-2-REL}_{\mathcal{X},\omega_{2,\mathcal{X}}}(\bar{x}^{2t},\bar{x}^2_{j_{1_{2,\mathcal{X}}}},\ldots, \bar{x}^2_{j_{|{\bar{\omega}_{2,\mathcal{X}}}|}}) \Rightarrow \]
\[
\exists \bar{x}^3(3\text{-REL}_{\mathcal{X}}(\bar{x}^3,\bar{x}^{2t}))\bigg] \bigg].
\]

\end{document}